\newcommand\mnras{MNRAS}

\newcommand\vr{\vec{r}}

\newcommand\vs{\vec{s}}
\newcommand\dO{d\Omega}
\newcommand\hr{\hat{r}}
\newcommand\de{\delta}

\documentclass[useAMS,usenatbib]{mn2e}

\usepackage{placeins}

\usepackage{graphicx} 
\usepackage{multirow}
\usepackage{array}

\usepackage{amsmath,amssymb}

\usepackage[T1]{fontenc}
\usepackage[latin9]{inputenc}
\usepackage{float}
\usepackage{graphicx}
\usepackage{esint}

\usepackage{hyperref}

\makeatletter
\DeclareFontEncoding{LGR}{}{}
\DeclareTextSymbol{\~}{LGR}{126}

\begin{document}

\title[An $\mathcal{O}(N^2)$ Three-Point Function]{Computing the Three-Point Correlation Function of Galaxies in $\mathcal{O}(N^2)$ Time}

\author{\makeatauthor}

\author[Slepian and Eisenstein]{Zachary Slepian\thanks{E-mail: zslepian@cfa.harvard.edu} and Daniel J. Eisenstein\thanks{E-mail: deisenstein@cfa.harvard.edu}\\
Harvard-Smithsonian Center for Astrophysics, Cambridge, MA 02138\\
}
\maketitle

\begin{abstract}
We present an algorithm that computes the multipole coefficients of the galaxy three-point correlation function (3PCF) without explicitly considering triplets of galaxies. Rather, centering on each galaxy in the survey, it expands the radially-binned density field in spherical harmonics and combines these to form the multipoles without ever requiring the relative angle between a pair about the central. This approach scales with number and number density in the same way as the two-point correlation function, allowing runtimes that are comparable, and 500 times faster than a naive triplet count. It is exact in angle and easily handles edge correction.  We demonstrate the algorithm on the LasDamas SDSS-DR7 mock catalogs, computing an edge corrected 3PCF out to $90\;{\rm Mpc}/h$ in under an hour on modest computing resources.  We expect this algorithm will render it possible to obtain the large-scale 3PCF for upcoming surveys such as Euclid, LSST, and DESI.
\end{abstract}


\section{Introduction}
\label{sec:intro}
In the current picture of structure formation, inflation ends in reheating, which produces Gaussian random field density fluctuations in the radiation, matter, and dark matter.  As a Gaussian random field, the density is described completely by its mean and 2-point correlation function (2PCF), which measures the probability of finding a certain value of the density at one point given the density at another.  However, the subsequent evolution of the density field introduces additional correlations as gravity drives the convergence of overdense regions towards each other.  In particular, a 3-point correlation function (3PCF) is produced by this evolution (Bernardeau et al. 2002 or Szapudi 2005 for reviews).  Since the evolution is itself sensitive to the cosmological parameters, measuring the 3PCF of galaxies offers an independent probe of these parameters. It is typically used to break the degeneracy between galaxy bias (encoding the fact that galaxies do not trace the matter density field with perfect fidelity) and the clustering on a given scale (e.g. $\sigma_8$) (Gazta{\~n}aga \& Frieman 1994; Jing \& Borner 2004; Guo et al. 2014).  The 3PCF measurements also can probe primordial non-gaussianity (Desjacques \& Seljak 2010); while the constraints on this are currently dominated by CMB experiments such as Planck, it is expected that the increasing quality and number of galaxy redshift surveys will provide interesting independent information.

Since the first measurement by Peebles \& Groth (1977), numerous studies have presented 3PCF measurements, summarized in Kayo et al. (2004), McBride et al. (2011a, b), Guo et al. (2014) and references therein.  In this work, we present a new algorithm for measuring the 3PCF of galaxies through its multipole moments.  This decomposition of the 3PCF was first advanced in Szapudi (2004) and to a limited extent (measurement of the monopole moment) used in Pan \& Szapudi (2005) on Two-degree-Field Galaxy Redshift Survey (2dFGRS) data. Slepian \& Eisenstein (2015) (hereafter SE15) found this decomposition to be particularly useful for distinguishing linear and non-linear bias as well as isolating a possible relative velocity bias.

Current algorithms, such as that used for the McBride et al. (2011) measurement (presented in Moore et al. 2001, Gray et al. 2004, Nichol et al. 2006, and Gardner et al. 2007) fundamentally scale as the number of possible triangles in a survey.  If one wishes to measure the 3PCF out to some scale $R_{\rm max}$, there are $N(nV_{R_{\rm max}})^2$ relevant triangles, where $N$ is the number of objects in the survey, $n$ is the survey number density and $V_{R_{\rm max}}=(4/3)\pi R_{\rm max}^3$.  The algorithm presented in the series of references above, whose most recent incarnation is developed in March (2013), uses multiple mrkd-trees. Here ``mr'' means the kd-tree caches additional information, in this case the number of galaxies within each node of the tree as well as the bounding box of the node.  This algorithm is faster than simply counting all triangles. It is particularly effective if the galaxies are close to each other, so that there are many triangles whose side lengths fall within a given combination of radial bins.

However, typical galaxy surveys are sparse, particularly those mapping the largest volumes. For example,  the Baryon Oscillation Spectroscopic Survey (BOSS) has an average separation of 13 ${\rm Mpc}/h$, too large to permit many galaxies to be in the same bin.  This means the algorithm will not be as fast for such large-scale measurements.  The use case tested in March (2013) is triangles with three sides of $8$ Mpc each, much smaller than the scales that are well-described by linear perturbation theory and hence most useful for cosmology.  Furthermore, even with the speed-ups coming from the multi-tree structure of the algorithm, it is still fundamentally scaling as the number of galaxies in the survey times the square of the number within $R_{\rm max}$ (March 2013,  Figure 21).  

 In this paper, we present an algorithm that does better: it scales as the number of galaxies in the survey times the number within $R_{\rm max}$,  and so by construction is significantly faster than any previous algorithm that is exact in angle. In brief, we write the opening angle dependence of the triangles about a given vertex in terms of Legendre polynomials of $\hat{r}_1\cdot\hat{r}_2$, where these are two unit vectors describing two triangle sides.  The dot product seems to require explicitly considering all pairs of galaxies about a given vertex (i.e. third galaxy), but using the spherical harmonic addition theorem, this representation can be factored into a product of spherical harmonics each depending on only one unit vector.  Therefore from the spherical harmonic expansion of the radially binned density field one can obtain the multipole moments without ever needing to consider pairs about a given vertex. This is the central insight of this paper.
 
In Section \ref{sec:algorithm}, we present the algorithm in more detail, and show in Section \ref{sec:proj3PCF} how this framework goes through to the projected 3PCF. Section \ref{sec:edgecorrxn} discusses edge correction, while Section \ref{sec:implementation} describes our implementation.  Section \ref{sec:covariance} computes the covariance of this multipole decomposition in the Gaussian random field limit, and Section \ref{sec:mock_results}  presents the results of using the algorithm on the LasDamas SDSS-II Data Release 7 (SDSS-DR7) Luminous Red Galaxy mock catalogs.  We conclude in Section \ref{sec:conclusions}.


\section{The algorithm}
\label{sec:algorithm}
\subsection{Legendre basis}
In this paper, we parametrize triangle configurations by two side lengths, $r_1$ and $r_2$, and the angle between them with cosine $\hr_1\cdot\hr_2$.  We will decompose the 3PCF as a function of these three variables into a sum over Legendre polynomials for the angular dependence times radial coefficients encoding the side length dependence, as
\begin{align}
\zeta(r_1,r_2;\hat{r}_1\cdot\hat{r}_2)=\sum\zeta_l(r_1,r_2)P_l(\hat{r}_1\cdot\hat{r}_2).
\label{eqn:zeta_series}
\end{align}
Szapudi (2004) first advanced this decomposition, and he puts a factor of $(2l+1)/(4\pi)$ in front of his analogous expansion coefficients; we absorb this into $\zeta_l$.  

There are three major advantages to this decomposition.  First, the shape of the 3PCF for fixed side lengths as a function of angle is smooth and slowly varying (see e.g. Bernardeau 2002, Figure 11), without much fine structure. Thus we expect that only a few multipoles will be required to capture the angle dependence.  Second, this decomposition provides a natural way to visualize the 3PCF for all triangle configurations; one can make several panels for different $l$, each with all $r_1$ and $r_2$ and amplitudes indicated by a colorbar, as in SE15.  In contrast to many previous works, this allows immediate appraisal of the information in all triangles and not just a particular set of configurations (e.g. isosceles, two-to-one, etc.)  

Third, as we will see,  the multipole moments of the 3PCF can be obtained with much greater speed than other decompositions of the 3PCF.  However, in contrast to other fast methods, such as tree methods that fix a critical angular scale below which they are approximate (e.g. Zhang \& Pen 2005) or Fourier methods that choose a grid with some minimum spacing, we do not sacrifice accuracy to obtain this speed. Our method is exact in angle. We will bin in side length, but even were speed of no concern this would be necessary to keep the covariance matrix to a reasonable size.

\subsection{Rotation and translation averaging}
\label{subsec:prelim}
The 3PCF describes the number of triangles of a given configuration whose vertices are the galaxies in a survey. While nine coordinates are required to completely describe any individual triangle connecting three galaxies, the 3PCF averages over both translations and rotations of the triangle configuration. The presumed losslessness of this averaging corresponds to the two usual cosmological assumptions of isotropy (rotation-invariance about a given point) and homogeneity (translation-invariance). This ultimately reduces the 3PCF to a function of only three variables; as indicated already, we will use two triangle sides and the angle between them. We will now show explicitly how to go from nine coordinates to three.

We begin with averaging over rotations.  We will show explicitly that Legendre polynomials are an angular basis for the 3PCF after this averaging.  To do so, we first step back and write an estimate (denoted by a hat) of the 3PCF for a triangle with sides $\vec{r}_1,\vec{r}_2$ extending from a vertex whose absolute position within the survey is $\vec{s}$.  We have
\begin{align}
\hat{\zeta}(\vec{r}_1,\vec{r}_2;\vec{s})=\sum_{lm}\sum_{l'm'} \hat{\zeta}_{ll'}^{mm'}(r_1,r_2;\vec{s})Y_{lm}(\hat{r}_1)Y_{l'm'}^*(\hat{r}_2).
\end{align}

We now wish to average over all rotations of the triangle about $\vec{s}$.  Writing a rotation as $\boldsymbol{R}$ (simply a matrix involving the three Euler angles), we have
\begin{align}
&\hat{\zeta}_{\rm iso}(r_1,r_2;\hat{r}_1\cdot\hat{r}_2;\vec{s})=\sum_{lm}\sum_{l'm'} \hat{\zeta}_{ll'}^{mm'}(r_1,r_2;\vec{s})\nonumber\\
&\times\int d\boldsymbol{R}\; Y_{lm}(\boldsymbol{R}\hat{r}_1)Y_{l'm'}^*(\boldsymbol{R}\hat{r}_2),
\end{align}
where subscript ``iso'' abbreviates ``isotropy.''
Noting that $Y_{lm}(\boldsymbol{R}\hat{r})=\sum_M D^l_{mM}Y_{lM}(\hat{r})$, where $D^l_{mM}$ is a Wigner matrix (e.g. Arfken, Weber \& Harris 2013 (hereafter AWH13), equation (16.52)), we find
\begin{align}
&\hat{\zeta}_{\rm iso}(r_1,r_2;\hat{r}_1\cdot\hat{r}_2;\vec{s})=\sum_{lm}\sum_{l'm'} \hat{\zeta}_{ll'}^{mm'}(r_1,r_2;\vec{s})\nonumber\\
&\times \sum_{MM'}Y_{lM}(\hat{r}_1)Y_{l'M'}^*(\hat{r}_2)
\int d\boldsymbol{R}\; D_{mM}^lD_{m'M'}^{l'*}.
\end{align}
The integral over Wigner matrices is simply evaluated by orthogonality (e.g. Brink \& Satchler 1993, Appendix V) as $8\pi^2/(2l+1)\delta^K_{ll'}\delta^K_{mm'}\delta^K_{MM'}$, $\delta^K$ the Kronecker delta.  Using the spherical harmonic addition theorem (AWH13, equation (16.57))
\begin{align}
P_l(\hat{r}_1\cdot\hat{r}_2)=\frac{4\pi}{2l+1}\sum_{m=-l}^l Y_{lm}(\hat{r}_1)Y_{lm}^*(\hat{r}_2),
\label{eqn:sph_addition_theorem}
\end{align}
and defining
\begin{align}
\hat{\zeta}_l(r_1,r_2;\hat{r}_1\cdot\hat{r}_2;\vec{s})=2\pi\sum_m \hat{\zeta}_{ll}^{mm}(r_1,r_2;\vec{s})
\end{align}
we find
\begin{align}
\hat{\zeta}_{\rm iso}(r_1,r_2;\hat{r}_1\cdot\hat{r}_2;\vec{s})=\sum_l \hat{\zeta}_l(r_1,r_2;\vec{s})P_l(\hat{r}_1\cdot\hat{r}_2).
\label{eqn:rotation_avg}
\end{align}
In what follows we drop the subscript ``iso'' as we will always be considering the isotropic 3PCF.

We now move to averaging over translations. Recalling that $\vec{s}$ is the vertex of the triangle from which the two sides given by $\vec{r}_1,\vec{r}_2$ extend, the densities on a particular triangle of points will be $\delta(\vs)\delta(\vec{r}_1+\vec{s})\delta(\vec{r}_2+\vec{s})$. Averaging over translations means allowing every point in the survey to serve as the vertex $\vec{s}$, so we must integrate over $d^3\vec{s}$. We thus find that the $l^{\rm th}$ radial coefficient of the 3PCF is
\begin{align}
\zeta_l(r_1,r_2)=\frac{1}{V}\int d^3\vec{s}\; \hat{\zeta}_l(r_1,r_2;\vec{s}),
\label{eqn:translation_avg}
\end{align}
where $V$ is the survey volume.

\subsection{Radial binning}
\label{subsec:binning}
Our algorithm will bin radially (denoted with a bar), so we seek
\begin{align}
\bar{\zeta}_l(r_1,r_2)=\int r^2 r'^2 dr dr' \zeta_{l}(r,r')\Phi(r;r_1)\Phi(r';r_2),
\label{eqn:binned_zeta}
\end{align}
with $\Phi$ a binning function demanding that we are in the bin given by its second argument. Binning averages the radial coefficient over some interval in each side length, and in that sense is not lossless.  It is also necessary for the speed advantage of our algorithm, as will become clear shortly. 

We will not compute using equation (\ref{eqn:binned_zeta}). Rather, we will bin radially around each possible origin $\vec{s}$ {\it before} averaging over rotations and translations, so it will be useful also to define the binned estimator before translation-averaging as
\begin{align}
&\bar{\hat{\zeta}}_l(r_1,r_2;\vec{s})=\nonumber\\
&\frac{2l+1}{(4\pi)^2}\int d\Omega_1 d\Omega_2 \delta(\vec{s})\bar{\delta}(r_1;\hat{r}_1;\vec{s})\bar{\delta}(r_2;\hat{r}_2;\vec{s})P_l(\hat{r}_1\cdot\hat{r}_2),
\label{eqn:binned_zetahat_l}
\end{align}
where
\begin{align}
\bar{\delta}(r_i;\hat{r}_i;\vec{s})=\int r^2 dr\; \Phi(r;r_i) \delta(\vec{r}_i+\vec{s})
\end{align}
is the radially binned density field about an origin $\vs$.

Hence in practice we never compute $\zeta_l(r,r')$ using equation (\ref{eqn:binned_zeta}), but rather measure $\bar{\hat{\zeta}}_l$ via equation (\ref{eqn:binned_zetahat_l}) and then compute
\begin{align}
\bar{\zeta}_l(r_1,r_2)=\frac{1}{V}\int d^3\vs\; \bar{\hat{\zeta}}_l(r_1,r_2;\vs)
\label{eqn:zeta_binned_final}
\end{align}
as the radially binned multipole coefficients of the 3PCF.
\subsection{Accelerating with spherical harmonics}
\label{subsec:legendredecomp}
A direct way to measure $\bar{\zeta}_l$ would be to sit on every possible origin and compute the angle between pairs of vectors pointing to all possible sets of two galaxies out to the radius $R_{\rm max}$ to which one wishes to measure the 3PCF.  This scales as $N(nV_{R_{\rm max}})^2$. As discussed in the Introduction, this scaling applies to other algorithms as well (e.g. the Gardner (2007) and March (2013) kd-tree approach), fundamentally because the number of possible triangles within $R_{\rm max}$ with one vertex fixed scales as $(nR_{\rm max})^2$.

\begin{figure}
\includegraphics[scale=0.4]{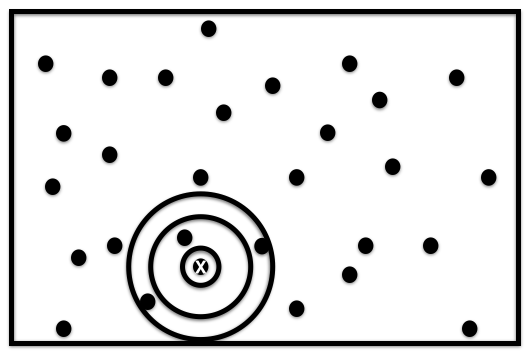}
\caption{Our algorithm sits on each galaxy in the survey, here marked with a white X, and computes the spherical harmonic expansion of the density field in concentric spherical shells (radial bins) around that point via equation (\ref{eqn:almsdef}). The $a_{lm}s$ can be combined to yield the multipole moments around this galaxy (sum over $m$, equation (\ref{eqn:zetal_ito_alms})) and then translation-averaged to yield $\zeta_l$ for the survey.}
\label{fig:kernel}
\end{figure}

However, as is the case for angular power spectra, we can exploit a property of multipole decompositions to enormously accelerate the measurement. We can use the spherical harmonic addition theorem (\ref{eqn:sph_addition_theorem}) to decompose the Legendre polynomial into factors that depend only on one angular variable each.  Inserting this into equation (\ref{eqn:binned_zetahat_l}), we find
\begin{align}
&\bar{\hat{\zeta}}_l(r_1,r_2;\vec{s})=\frac{1}{4\pi}\delta(\vec{s})\sum_{m=-l}^l \int \dO_1 \; \bar{\de}(r_1;\hr_1;\vs)Y_{lm}(\hr_1) \nonumber\\
&\times \int \dO_2 \;\bar{\de}(r_2;\hr_2;\vs)Y^*_{lm}(\hr_2).
\label{eqn:zetal_int}
\end{align}
This equation immediately shows how to reduce the quadratic scaling in the number density to a linear scaling.  The two angular integrals have now been separated, and each simply asks for a particular expansion coefficient of the density field (as a function of angle alone) in spherical harmonics, in a fixed radial bin.  In other words, if we compute for each radial bin $r$
\begin{align}
a_{lm}(r;\vec{s})&\equiv \int \dO \; \bar{\de}(r;\hr; \vs)Y^*_{lm}(\hr)\nonumber\\
&=\int \dO \;Y_{lm}^*(\hat{r})\int r'^2 dr' \Phi(r';r)\delta(\vr'+\vs)
\label{eqn:almsdef}
\end{align}
we can construct all combinations dictated by $r_1$ and $r_2$, without ever needing to do an $\mathcal{O}(n^2)$ operation. Explicitly, inserting equation (\ref{eqn:almsdef}) into equation (\ref{eqn:zetal_int}), we find
\begin{align}
&\bar{\hat{\zeta}}_l(r_1,r_2;\vec{s})=\frac{1}{4\pi}\delta(\vec{s})\sum_{m=-l}^l a_{lm}(r_1;\vs)a_{lm}^*(r_2;\vs).
\label{eqn:zetal_ito_alms}
\end{align}
This is why radial binning is essential for the speed-up of our algorithm; we can precompute the $a_{lm}(r;\vs)$s in each radial bin. For $N_{\rm bins}$, we then only need to construct $(N_{\rm bins}+1)N_{\rm bins}/2$ combinations of these coefficients. A schematic about a single possible origin is shown in Figure \ref{fig:kernel}.

For a 3PCF measurement, one might use a bin width $\sim 10$ Mpc, and so if one measures out to $200\;{\rm Mpc}$ there will be only $210$ distinct bin combinations. Meanwhile, computing the $a_{lm}(r;\vs)$s themselves takes only as long as performing the integral (\ref{eqn:almsdef}), which should scale as $nV_{R_{\rm max}}$.

We still must integrate over all possible choices of origin as dictated by equation (\ref{eqn:translation_avg}).  Because the galaxies are discrete, this will reduce to a sum with $N$ terms. Thus our algorithm will scale as $N(nV_{R_{\rm max}})$: linear in both the total number of galaxies {\it and} the number within a sphere of radius $R_{\rm max}$, and a factor of order $(nV_{R_{\rm max}})$ faster than the naive counting approach. Our algorithm thus provides a route to the 3PCF that on large scales is no more computationally intensive than calculating the multipole moments (standardly calculated are monopole and quadrupole) of the 2-point correlation function (2PCF).

Finally, to obtain the spherical harmonic coefficients of the galaxy density as in equation (\ref{eqn:almsdef}), one might think a spherical harmonic transform is required. This scales as $N_{\rm g}^{3/2}$, $N_{\rm g}$ the number of spatial grid cells on the surface of a sphere. A large number of grid cells is necessary for accuracy even if there are very few galaxies, much as a small $\Delta k$ is needed when taking a numerical Fourier transform to avoid ringing. However, because only low-order multipoles are needed here ($l \lesssim 10$), we can avoid this transform and instead directly evaluate the $a_{lm}$s, which are simply spherical harmonics evaluated at angles given by a galaxy's location with respect to a given choice of origin. The required $Y_{lm}$s can be easily computed using the Cartesian expressions for the spherical harmonics (e.g. AWH13, equation (15.139) and Table 15.4). Indeed, about a given origin, the Cartesian components $x/r, y/r,$ and $z/r$ and their powers for each galaxy can be pre-calculated just once and subsequently combined to form all of the required multipoles. 
\section{Projected 3PCF}
\label{sec:proj3PCF}
 Redshift-space distortions (RSD) are differences between the true position of a galaxy along the line of sight and its position as inferred from assuming its redshift is purely cosmological.  They arise from peculiar velocities, ultimately generated by the growth of large-scale structure (Hamilton 1998, for a review).  The projected 3PCF is insensitive to these distortions because it is integrated along the line of sight. Below we show how our approach extends to measuring it.

We work in the flat-sky approximation, where there is a single line of sight to all galaxies in the survey. Sitting around a given central galaxy and projecting corresponds to drawing cylindrical shells around that central with bases that are concentric annuli. All of the galaxies in a given cylinder project down into the cylinder's base annulus.  We thus have a planar problem with circular symmetry.  

This permits simplification of our spherical harmonic basis. Recall that
\begin{align}
Y_{lm}(\theta,\phi)=\sqrt{\frac{2l+1}{4\pi}\frac{(l-m)!}{(l+m)!}}P_{lm}(\cos\theta)e^{im\phi},
\end{align}
where here $\theta,\phi$ are the angular coordinates of a galaxy in the system where the central is at the origin. Since all the (projected) positions are coplanar with the central, the separation along the $z$-axis is zero, so $\cos\theta=0$. Defining 
\begin{align}
b_{lm}=\sqrt{\frac{2l+1}{4\pi}\frac{(l-m)!}{(l+m)!}}P_{lm}(0),
\end{align}
we see from equation (\ref{eqn:zetal_int}) that the multipole moments of the projected, radially binned 3PCF will simply involve Fourier coefficients of the projected, radially binned density field weighted by $b_{lm}$:
\begin{align}
&\bar{\hat{\zeta}}_{l,{\rm proj}}(r_1,r_2;\vec{s})=\frac{1}{4\pi}\nonumber\\
&\times \sum_{m=-l}^l b_{lm} \int d\phi_1 \bar{\delta}(r_1;\phi_1;\vec{s})e^{im\phi_1} \int d\phi_2 \bar{\delta}(r_2;\phi_2;\vec{s})e^{-im\phi_2}.
\end{align}
Above, the integrals over $\theta_1$ and $\theta_2$ of equation (\ref{eqn:zetal_int}) have already been performed using that the projected density field is only non-zero at $\theta=\pi/2$.

With this in mind, we observe that if one is solely interested in the projected 3PCF, it is probably optimal simply to use the Fourier basis directly.  One parametrizes the projected 3PCF estimator about a given central as 
\begin{align}
\bar{\hat{\zeta}}_{\rm proj}(r_1,r_2;\theta_{12};\vec{s})=\sum_m \bar{\hat{\zeta}}_{{\rm proj},m}(r_1,r_2;\vs)e^{i m\theta_{12}}
\label{eqn:fourier_param}
\end{align}
and writes the exponential as
\begin{align}
e^{i m\theta_{12}}=\left(e^{im \theta_2}\right)\left(e^{im\theta_1}\right)^*
\end{align}
where $\theta_1$ and $\theta_2$ are now angles in the plane in polar coordinates, with $\theta_{12}=\theta_2-\theta_1$. Using orthogonality of the plane waves, one may then extract the expansion coefficients $\bar{\hat{\zeta}}_{{\rm proj},m}(r_1,r_2;\vs)$ in equation (\ref{eqn:fourier_param}) as \begin{align}
&\bar{\hat{\zeta}}_{{\rm proj},m}(r_1,r_2;\vs)=\nonumber\\
&\frac{\delta(\vec{s})}{(2\pi)^2}\int d\theta_1 \;\bar{\delta}(r_1;\theta_1;\vec{s})e^{im\theta_1}\int d\theta_2 \; \bar{\delta}(r_2;\theta_2;\vec{s})e^{-im\theta_2}.
 \end{align}
Just as in the non-projected case, these integrals can be explicitly evaluated using the Cartesian expressions for the exponentials, and precomputing $x/r$ and $y/r$. Again, one never explicitly considers pairs of galaxies about a given central; one simply constructs the coefficients 
\begin{align}
c_m(r;\vec{s})\equiv \int d\theta \; \bar{\delta}(r;\theta;\vec{s})e^{im\theta}
\end{align}
 for all radial bins, then computes 
 \begin{align}
 \bar{\hat{\zeta}}_{{\rm proj},m}(r_1,r_2;\vs)=\delta(\vs)c_m(r_1;\vs)c_m^*(r_2\vs)/(2\pi)^2
 \end{align}
  for all desired bin combinations, and finally averages over translations by integrating out $\vs$. We should note that Chen \& Szapudi (2005) advanced a similar scheme to measure the 3PCF of Cosmic Microwave Background (CMB) maps, analogous to the projected 3PCF since both are on 2-D manifolds.  However their method evaluates the Fourier transform of the (continuous) temperature anisotropy map by gridding, whereas here we suggest the (discrete) galaxy density field be Fourier-transformed using direct evaluation of the Cartesian expressions for $x/r$ and $y/r$.

\section{Edge correction}
\label{sec:edgecorrxn}
Surveys have jagged and complicated boundaries, and these can produce a spurious contribution to the 3PCF that is the signature of the survey geometry rather than physics the survey hopes to probe.  This spurious contribution must be removed.  In Fourier space, boundaries lead to Gibbs phenomenon ringing in the bispectrum, and are challenging to remove. However, in configuration space, edge correction is fairly straightforward for popular estimators (see Kayo et al. 2004, Appendix, for comparison of several).  

We focus here on the Szapudi \& Szalay (1998) estimator, which Kayo et al. (2004) find preferable to the others they consider; it has now become the standard in the field. It is
\begin{equation}
\hat{\zeta}=\frac{NNN}{RRR}
\label{eqn:hat_zeta_est}
\end{equation}
with $N\equiv (D-R)$, $D$ the data and $R$ the random counts. Note that, if one inserted $N/R$ for $\delta$ in Section \ref{sec:algorithm}, one would need to compute integrals of this fraction against the spherical harmonics, requiring definition of $N/R$ at every point in space.  However, the estimator (\ref{eqn:hat_zeta_est}) really represents the function
\begin{align}
\zeta_{\rm raw}(\vec{r}_1,\vec{r}_2,\vec{r}_3)=\frac{N(\vec{r}_1)N(\vec{r}_2)N(\vec{r}_3)}{R(\vec{r}_1)R(\vec{r}_2)R(\vec{r}_3)}
\label{eqn:unwted_estimator}
\end{align}
averaged over rotations and translations with weights $w=R(\vec{r}_1)R(\vec{r}_2)R(\vec{r}_3)\theta$, which in the shot noise limit is just inverse variance weighting (we include radial binning represented by $\theta$). In short,
\begin{align}
&\hat{\zeta}=\frac{\int d^3\vec{r}_1 d^3\vec{r}_2d^3\vec{r}_3\; w(\vec{r}_1,\vec{r}_2,\vec{r}_3) \zeta_{\rm raw}(\vec{r}_1,\vec{r}_2,\vec{r}_3)}{\int d^3\vec{r}_1 d^3\vec{r}_2d^3\vec{r}_3\; w(\vec{r}_1,\vec{r}_2,\vec{r}_3)} \nonumber\\
&= \frac{\int d^3\vec{r}_1 d^3 \vec{r}_2 d^3\vec{r}_3\; \theta NNN}{\int d^3\vec{r}_1 d^3 \vec{r}_2 d^3 \vec{r}_3 \;\theta RRR}.
\end{align}
Thus the estimator (\ref{eqn:hat_zeta_est}) should be interpreted as demanding the triple count $NNN$ divided by the triple count $RRR$. Therefore we can insert $N$ and $R$ separately in turn for $\delta$ in Section \ref{sec:algorithm}, processing random and data catalogs serially.  The division required can be done as a post-processing step.  We now turn to how this division translates to the Legendre basis.

\subsection{Edge correction in the Legendre basis}
Working now in our Legendre basis, we have
\begin{equation}
\hat{\zeta}=\sum_l \hat{\zeta}_{l}(r_{1},r_{2})P_{l}(\hat{r}_{1}\cdot\hat{r}_{2})
\label{eqn:mp_NR}
\end{equation}
\begin{equation}
NNN=\sum_j \mathcal{N}_{j}(r_{1},r_{2})P_{j}(\hat{r}_{1}\cdot\hat{r}_{2})
\label{eqn:mp_N}
\end{equation}
and
\begin{equation}
RRR=\sum_{l'}\mathcal{R}_{l'}(r_{1},r_{2})P_{l'}(\hat{r}_{1}\cdot\hat{r}_{2}).
\label{eqn:mp_R}
\end{equation}
Inserting the multipole expansions (\ref{eqn:mp_NR})-(\ref{eqn:mp_R}) into the estimator (\ref{eqn:hat_zeta_est}) and multiplying through by $RRR$ we find
\begin{equation}
\sum_{ll'}\mathcal{R}_{l'}\hat{\zeta}_{l}P_{l'}(\hat{r}_{1}\cdot\hat{r}_{2})P_{l}(\hat{r}_{1}\cdot\hat{r}_{2})=\sum_{j}\mathcal{N}_{j}P_{j}(\hat{r}_{1}\cdot\hat{r}_{2}).
\end{equation}

Using a linearization formula for the product of two Legendre polynomials  (Ferrers (1877), Adams (1878), Neumann (1878), Park \& Kim (2006); SE15 equation (A11)) we find, with angular arguments suppressed,
\begin{equation}
\sum_{l'lj'}\mathcal{R}_{l'}\hat{\zeta}_{l}(2j'+1)\left(\begin{array}{ccc}
l & l' & j'\\
0 & 0 & 0
\end{array}\right)^{2}P_{j'}=\sum_{j}\mathcal{N}_{j}P_{j}.
\end{equation}
The Wigner 3j-symbol above describes angular momentum coupling; see e.g. Brink \& Satchler (1993) or AWH13.  The vector addition of angular momenta means that the upper row must satisfy triangle inequalities, so $|l-l'|\leq j'\leq l+l'$ and at fixed $l$ and $l'$ the sum is finite.  Using orthogonality, separating out the $l'=0$ term, dividing through by $\mathcal{R}_0$, and defining $f_{l'}=\mathcal{R}_{l'}/\mathcal{R}_{0}$, we obtain
\begin{equation}
\frac{\mathcal{N}_{k}}{\mathcal{R}_{0}}=\hat{\zeta}_{k}+\sum_{l}\hat{\zeta}_{l}(2k+1)\sum_{l'>0}\left(\begin{array}{ccc}
l & l' & k\\
0 & 0 & 0
\end{array}\right)^{2}f_{l'}.
\label{eqn:edgecorrxn_fund}
\end{equation}

For a boundary-free survey the random field would generate
only a monopole ($\mathcal{R}_0$), leaving only $\hat{\zeta}_k$ on the righthand side; this is the limit where there is no need for edge-correction, but just division by the randoms. The form of equation (\ref{eqn:edgecorrxn_fund}) suggests
that this problem can be cast as a matrix multiplication, so we define the multipole coupling
matrix $\boldsymbol{M}$ with elements
\begin{equation}
M_{kl}=(2k+1)\sum_{l'>0}\left(\begin{array}{ccc}
l & l' & k\\
0 & 0 & 0
\end{array}\right)^{2}f_{l'}.
\label{eqn:Mkl}
\end{equation}
Note that while these matrix elements describe the off-diagonal couplings of different multipoles to each other, they need not be zero along the diagonal. A given multipole in the data may couple to that same multipole in $\hat{\zeta}$ because the 3j-symbol allows $l=k$ for $l'>0$.  But the dominant coupling of a given multipole in the data to the same multipole in $\hat{\zeta}$ is described by $\hat{\zeta}_k$ in equation (\ref{eqn:edgecorrxn_fund}), since the $f_{l'}$ are expected to be much less than unity.  This term translates to the identity matrix $\boldsymbol{I}$.  The edge-correction equation (\ref{eqn:edgecorrxn_fund}) thus becomes
\begin{equation}
\vec{\mathcal{N}}/\mathcal{R}_{0}=(\boldsymbol{I}+\boldsymbol{M})\vec{\hat{\zeta}}\equiv\boldsymbol{A}\vec{\hat{\zeta}},
\label{eqn:matrix_edgecorrxn}
\end{equation}
where $\vec{\mathcal{N}}=(\mathcal{N}_{0},\mathcal{N}_{1},\cdots,\mathcal{N}_{l_{\rm max}})$ and analogously
for $\vec{\hat{\zeta}}$. The system of equations this represents can then be
solved for $\vec{\hat{\zeta}}$ by matrix inversion. 

To explore this matrix for a realistic use case, we use the LasDamas SDSS DR7 real space mock catalogs, using 15 radial bins and a maximum scale of $90\;{\rm Mpc}/h$ (further details are given in Section \ref{sec:mock_results}).  We show $M_{kl}$ for a particular bin in $(r_1,r_2)$ in Figure \ref{fig:Mkl}, and the leading order edge correction factor $f_1$ in Figure \ref{fig:f_one}. $\boldsymbol{M}$ is not symmetric, but $M_{kl}/(2k+1)$ is; this is why the upper off-diagonal, where $k>l$, exceeds the lower in Figure \ref{fig:Mkl}.  

\subsection{Solving the edge correction equation}

There are two approximations implicit in our approach to solving equation (\ref{eqn:matrix_edgecorrxn}).  First, to obtain a given matrix element $M_{kl}$, formally one requires $f_{l'}$ for all values of $l'$.  However, for $l'>1$ these factors fall rapidly.  For the LasDamas real space mock catalogs for which we present results here, they are $\lesssim 0.1\%$ by $l'=10$ even for the largest-scale radial bin combination (the values are listed in the caption to Figure \ref{fig:Mkl}), so we simply truncate the series there.  If one wished one could easily expand our code to measure higher multipoles of the randoms at the cost of slightly more computation time. However we expect that going to $l'=10$ will already render the edge correction error negligible compared to the total error budget.  

Importantly, the smallness of the $f_{l'}$ for $l'>1$ means that the coupling between multipoles $k$ and $l$ is nearly diagonal. Coupling between multipoles separated by more than one angular momentum step is suppressed as $f_2$ or higher because the 3j-symbol in the coupling matrix elements (\ref{eqn:Mkl}) requires that $l'>|l-k|$.  


The second approximation relates to the matrix inversion when we solve equation (\ref{eqn:matrix_edgecorrxn}).  Formally one has an infinite dimensional matrix where at fixed $l$, all $k$ enter the correction.  Thus this matrix will not be square (and hence invertible) unless we go to an infinite number of $l$ as well.  However, in practice the matrix is so diagonally-dominant that we believe it is accurate enough simply to invert the sub-matrix given by truncating $l$ and $k$ at some maximum multipole.  We verify this approximation by constructing $M_{kl}$ using solely the dominant $f_1$ edge-correction factor, letting $k$ and $l$ go to $2l_{\rm max}$, inverting, and comparing to the result where both go to $l_{\rm max}$. Were the matrix purely diagonal, truncation would not affect the inverse at all. In the limit where only $f_1$ is non-zero (in reality, it does dominate the other edge correction factors), the matrix is tridiagonal, and so truncation at $l_{\rm max}$ affects $\hat{\zeta}_{l_{\rm max}}$ at order $f_1$, $\hat{\zeta}_{l_{\rm max}-1}$ at order $f_1^2$, and $\hat{\zeta}_{l_{\rm max}-n}$ at order $f_1^{n+1}$.

\subsection{A model for the edge correction factors}
Using a simple toy model, we can estimate the edge correction factors $f_l$ to confirm that they really should be small.  Consider a spherical ball of random galaxies with radius $R$ about a given central, and assume this sphere is cut by a planar survey boundary.  Orient the $z$-axis perpendicular to this boundary, with the central galaxy a distance $z$ from it. The problem now has symmetry about this axis, so we need only compute the $m=0$ spherical harmonic coefficients $a_{lm}$; $Y_{l0}=\sqrt{(2l+1)/4\pi}P_l(\mu)$, with $\mu=\cos\theta$. For a galaxy at distance $R$ from the central, there will be some critical angle with cosine $\mu_{\rm c}=z/R$ such that, for smaller $\mu$, the galaxy is outside the survey.  We have
\begin{align}
&a_{l0}=2\pi\sqrt{\frac{2l+1}{4\pi}} \int_{-1}^{\mu_{\rm c}} d\mu P_l(\mu)\nonumber\\
 &=\frac{2\pi}{2l+1}\sqrt{\frac{2l+1}{4\pi}} \left[ P_{l+1}(\mu_{\rm c})-P_{l-1}(\mu_{\rm c})\right].
 \end{align}
We used the recursion formula $(2n+1)P_n(\mu)=d/d\mu \left[P_{n+1}(\mu)-P_{n-1}(\mu)\right]$ to evaluate the integral and noted that the terms at the lower bound cancel off because they have the same parity. We now compute the $\hat{\zeta}_l=a_{l0}^2/(4\pi)$ required by equations (\ref{eqn:almsdef}) and (\ref{eqn:zetal_int}) and average over $\mu_{\rm c}$ (denoted by angle brackets).  We have
\begin{align}
&\left< \hat{\zeta}_l\right>=\frac{1}{4(2l+1)}\int _0^1 d\mu_{\rm c} \big[P_{l+1}^2(\mu_{\rm c})\nonumber\\
&-2P_{l+1}(\mu_{\rm c})P_{l-1}(\mu_{\rm c})+P_{l-1}^2(\mu_{\rm c})\big].
\end{align}    
Since each term above has even parity, we can integrate from $-1$ to $1$, divide by $2$, and then invoke orthogonality, to find that
\begin{align}
\left< \hat{\zeta}_l\right>=\frac{1}{2(2l+3)(2l-1)}.
\end{align}
Finally, we compute $\left< \hat{\zeta}_0\right>=7/12$ explicitly, to find that
\begin{align}
f_l\equiv \frac{\left< \hat{\zeta}_l\right>}{\left< \hat{\zeta}_0\right>}=\frac{6}{7}\frac{1}{(2l+3)(2l-1)}.
\label{eqn:fl_random_boundary}
\end{align}
$f_1=17.14\%$, $f_2=4.08\%$, $f_3=1.90\%$, $f_4=1.11\%$, falling to $f_{10}=0.196\%$. It should be kept in mind that in a large survey volume such as SDSS, many centrals will have spheres around them that do not impinge on a large-scale survey boundary at all, further reducing these factors; for instance, for the SDSS BOSS DR10 footprint only of order $20\%$ of spheres impinge on a boundary, so our rough estimates should be scaled down by a factor of 5. On the other hand, the true survey mask is far more complicated than the simple planar boundary model above, so this model should not be taken too literally.

\begin{figure}
\includegraphics[scale=0.45]{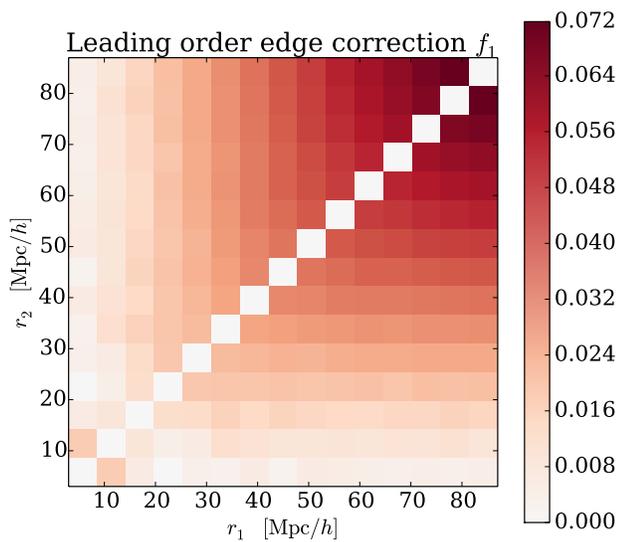}
\caption{The leading-order edge correction factor in equation (\ref{eqn:edgecorrxn_fund}) for the LasDamas SDSS DR7 real space mock catalogs with 15 radial bins out to $90\;{\rm Mpc}/h$. Higher $f_{l'}=\mathcal{R}_{l'}/\mathcal{R}_0$ fall off very rapidly.  Even the leading order coefficient is small. This means that one does not need to measure many multipoles of the randoms to obtain a highly accurate edge correction: since the higher $f_{l'}$ fall off so rapidly they contribute very little to the matrix $\boldsymbol{M}$ that must be inverted (equation (\ref{eqn:matrix_edgecorrxn})). As we expect, $f_1$ becomes larger at larger scales, as larger scale triangles are more likely to impinge on a survey boundary.}
\label{fig:f_one}
\end{figure}

\begin{figure}
\includegraphics[scale=0.45]{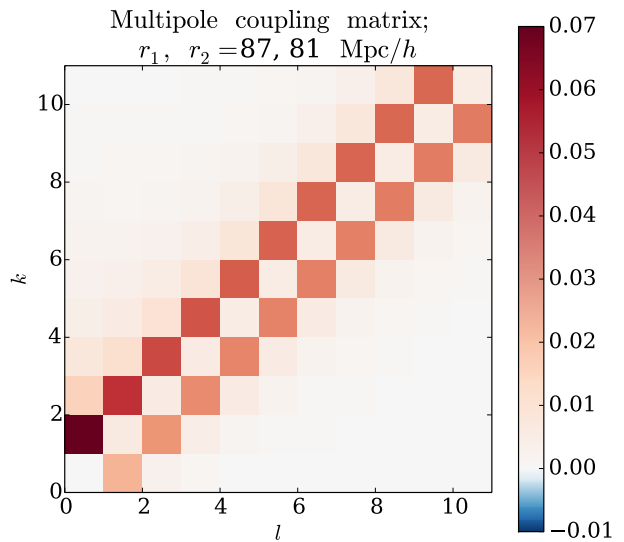}
\caption{The multipole coupling matrix elements (\ref{eqn:Mkl}) at each $k$ and $l$ for the largest combination of radial bins we test here. This illustrates that all of the couplings are $\lesssim 10\%$, even for the largest scales we test, which should have the largest correction factors as they are most likely to impinge on a survey boundary (see Figure \ref{fig:f_one}). While the diagonal appears zero in this plot, it is actually just small, as we discuss in the main text. The $f_l$ entering the matrix elements $M_{kl}$ for this radial bin combination are $f_1=7.18\%,\; f_2=1.59\%, \; f_3=0.783\%, \; f_4=0.428\%,\; f_5=0.318\%,\; f_6= 0.22\%, \; f_7=0.166\%,\; f_8 = 0.078\%,\; f_9=0.078\%$, and $\;f_{10}=0.051\%.$}
\label{fig:Mkl}
\end{figure}

\section{Implementation}
\label{sec:implementation}
\newcommand{\Rmax}{\ensuremath{R_{\rm max}}}
\newcommand{\hMpc}{\ensuremath{h^{-1}\ {\rm Mpc}}}

We next describe our C++ implementation of the ideas in Sections \ref{sec:algorithm} and \ref{sec:edgecorrxn}.  The basic
program flow is to loop over each central galaxy.  For each, we
find all neighbors within \Rmax\ and accumulate the $a_{lm}$ for
each of the radial bins.  Once finished with the neighbor finding,
we compute all of the bin cross-powers and add them to our accumulators
as a function of bins $r_1$ and $r_2$ and multipole $l$.  All of the
accumulations include a user-supplied weight per galaxy.

We accelerate the finding of neighbors by sorting the particles
into a grid, so that the search for neighbors need only consider
grid cells that include some point closer than $\Rmax$.  Ideally
one wants the grid spacing to be a few times smaller than $\Rmax$,
so that the inefficiency of doing a cubic search for a spherical
region is mild.  One also wants the grid spacing to be large enough
to contain at least several particles, so that the overhead of
storing and accessing the grid is modest.  These criteria are not
hard to satisfy: for the LasDamas mocks, 
we use a grid spacing of $50\;{\rm Mpc}/h$ when searching out to 
$\Rmax = 90\;{\rm Mpc}/h$; this typically contains
a dozen galaxies (and somewhat more random points).  For the SDSS-III Baryon
Oscillation Spectroscopic Survey, the density is three times
higher.

Once a neighbor is found, we need to add its contribution to the
spherical harmonics.  We do not use an angular binning to compute
the spherical harmonics. Rather, as mentioned in Section \ref{sec:algorithm}, we use the fact that the spherical
harmonics can be written as powers of the Cartesian coordinates of
unit vectors.  In particular, for a unit vector $\hat{r} = (x,y,z)$,
we can write $Y_{lm}(\hat{r})$ as a polynomial of terms of the form
$x^i y^j z^k$ where $i+j+k\le l$.  To compute all $a_{lm}$ up to
multipole order $p$, we therefore accumulate sums over all neighbors
of the Cartesian powers $x^i y^j z^k$ with $i+j+k\le p$, using the
unit vectors of the separation of the neighbor from the central
galaxy.  There are $(p+1)(p+2)(p+3)/6$ such power combinations for
each radial bin.  Having finished with all neighbors, we convert
these powers into the $a_{lm}$ using the appropriate coefficients
from the spherical harmonics, then form all of the bin-to-bin cross
powers.

For values of $p$ of order 10, the computation of the Cartesian
powers is much faster than doing the spherical harmonic transform
of a fine angular grid.  This is particularly true because we use
custom assembly code, supplied by Marc Metchnik as part of the
Abacus project (Metchnik \& Pinto, in prep.), to accumulate these
powers using Advanced Vector Extension (AVX) instructions.  In double precision, 8 neighbors
are computed at once, using two sets of AVX registers.

Though we do not present the 3PCF measurement here, we also run our algorithm on the SDSS-III BOSS DR10 data. In the North Galactic Cap footprint for the CMASS
sample, we consider the RRR count of 642,619 random particles.  We
count 6.7 billion pairs with $\Rmax=200\;{\rm Mpc}/h$, an average of 10,400
neighbors per central, divided into 10 linearly spaced radial bins.
Using $p=10$, the code runs in 170 seconds on a 6-core 4.2 GHz i7-3930K.  If we use $p=0$, thereby reducing the
problem to the pair finding and a simple accumulation per radial
bin, then the code runs in 53 seconds.  Loading the particles and
sorting them into the grid is a small fraction of that total, so
we infer that each pair found and processed at $p=0$ takes about
200 clock cycles.  Given $p=10$, we have 286 powers to track per
neighbor, each requiring a separate multiply and add.  Hence, we
are computing about 3.8 trillion double-precision operations in 120
extra seconds, a rate of 32 double-precision GFLOPS. 
 This is about 30\% of the maximum performance of the CPU (assuming 4 double
precision operations in AVX per clock cycle per core), a high mark
for a practical calculation.  The code is sustaining 22 GFLOPS for
the full problem, including the pair finding.

A two-point correlation function code would only need to count half
as many pairs, since the particles are indistinguishable in that application, and so at these speeds would take of order $53/2=27$ seconds.
We therefore find that our computation of the three-point correlation function
up to $l=10$ is only about $170/27\approx 6$ times slower than
the equivalent two-point correlation function calculation.
We would expect an explicit counting of triples to be about 3,000 times slower than the two-point pair counting, given the 10,000
neighbors (divide by a factor of 3 for the number of indistinguishable triples compared
to indistinguishable pairs).  As our method is only six times slower than a two-point measurement, it is
a factor of five hundred faster than an explicit triple count for this large-scale example.

In any method that compares the data points to a random set, we have
to consider the effect of Poisson noise in the randoms.  For example, 
in the Landy-Szalay (1993) estimator for the two-point function, $\hat{\xi}=(DD-2DR-RR)/RR$, we will have
noise in the data-random $(DR)$ and random-random $(RR)$ counts that would go to
zero in the limit of infinite numbers of random points.  One therefore
usually wants to use many more randoms than data (but note the important
optimization presented in Padmanabhan et al.\ (2007) in which one fits these 
counts to smooth functions of scale so as to reduce the Poisson noise).
A common inefficiency, however, is to use the same number of randoms for
each of the terms.  This results in spending far too much computational
resource on $RR$, whose Poisson noise will be dwarfed by the $DR$ noise.
For example, if the number of randoms is $m$ times the number of data points,
then (assuming uniform galaxy weights) the variance on $RR$ will be $1/m^2$ 
of that of $DD$ since the number of $RR$ pairs is $m^2$ the number of $DD$ pairs. In contrast, the variance of $DR$ will only be reduced to $(1/2)(4/m)$ of that of $DD$; the factor of 4 comes
from the factor of 2 in the Landy-Szalay estimator, and the $1/2$ enters because the $DD$ and $RR$ pairs have double the variance since each pair is counted twice.  Meanwhile the work 
in the two terms is scaling as $m^2/2$ and $m$, respectively.  

A simple way to avoid this is to compute the $DR$ and $RR$ counts with a 
smaller set of randoms and then repeat this numerous times, averaging
over the answers.  By choosing the number of randoms in each set, one 
can optimize the work.  For example, in the above two-point case,
at fixed total work, the number of random catalogs $N_{\rm cat}$ one can use scales as $(m^2/2+m)^{-1}$.  The total variance scales as the variance per random catalog divided by $N_{\rm cat}$, so as $(2/m+1/m^2)(m^2/2+m)$.
This is minimized for $m=1$, i.e., it is optimal to use random catalogs equal in size to the data set. A further advantage of this
method is that in addition to averaging all of the sets to get the best
answer, one can compute the variance to explicitly measure the
contribution of the random catalog density relative to one's estimate
of the irreducible on-sky variance.

For our three-point algorithm, the work scales as $2m+m^2$, while the
Poisson variance of $(D-R)^3$ for each random catalog scales as $1/m^3$ for $RRR$ and $3^2/m^2\times (2/6)$ for $DRR$ and $DDR$; the $3^2$ enters due to the $3$ in the Szapudi-Szalay estimator, while the $2/6$ comes from a 6-fold counting symmetry in $DDD$ and $RRR$ compared to a 2-fold one in $DDR$ and $DRR$. The total variance is thus $3/m+3/m^2+1/m^{3}$. At fixed total work $N_{\rm cat}$ scales as $(2m+m^2)^{-1}$, and so the total variance scales as $(3/m+3/m^2+1/m^3)(2m+m^2)$. This is minimized for $m=1.76$ but with only $1\%$ variation between $m=1.5$
and $2$.

We implement this strategy in our three-point method by supplying
a single list of particles, with the randoms concatenated to the
data but with negative weights.  Notationally, this is $N=D-R$, as in Section \ref{sec:edgecorrxn}.  We
then compute the three-point correlations of this $N$ list.  We
then re-run repeatedly with new random points $R$.  We avoid the
small amount of repeated counting of the $DD$ pairs and $DDD$ triples
by the following trick.  We first run the code with only the data
particle list and save a file that contains the Cartesian multipoles
for each radial bin and each primary particle, in the enumerated
order of the particles.  When next running with $D-R$ lists, whenever
a data particle is the primary (as marked by its having a non-negative weight), 
we initialize the multipole
accumulators with the saved values and then skip any secondary
particles that are also from the data list.  The resulting sums pass
transparently to the rest of the analysis code.  

We also run a separate case with only the randoms, so that we can
compute the denominator and edge-correction terms in equation (\ref{eqn:edgecorrxn_fund}).  This requires
much less precision, as the denominator of the estimator is much larger
than the numerator for large-scale correlations.  We therefore do this 
with only a single set of random points.

Finally, we have also written a Python implementation of the algorithm presented here and tested it on a periodic box with sides of $400\; {\rm Mpc}/h$ containing $20,000$ galaxies (roughly the SDSS BOSS number density). Rather than using gridding, this code exploits kd-trees for galaxy finding, using a fast C implementation (wrapped to python) in the $spatial$ library within $scipy$. We verified the accuracy of this code on a sample of 500 galaxies by comparing with a simple direct-counting algorithm that just counts triplets and then projects onto multipoles. This provides an important cross check on our spherical harmonics since the simple triple counting never uses spherical harmonics. We then ran both the multipole Python code and the multipole C++ code on a larger, 20,000 galaxy sample to verify the C++ code. Runtime for the Python version on a dual core (2014) MacBook Air was about $30$ minutes; since the box is periodic, scaling to larger numbers of galaxies is linear.

\section{Covariance matrix}
\label{sec:covariance}
Parameter fitting requires weighting the data points according to how independent they are, with two highly independent points contributing more than two less independent points all else equal. The covariance matrix describes how independent the measured multipoles at each $(r_1,r_2)$ are. For our algorithm to be useful, we must show that the covariance matrix can be controlled; here we compute it with this end in mind. The general 3PCF covariance has been computed before (Szapudi 2001) as a 6-D integral, but it is not straightforward to obtain the covariance of our multipole decomposition from this result. Here we derive the covariance for our multipole decomposition and show that it can be reduced to a sum of 2-D integrals. This reduction offers a significant improvement in the computation speed possible at a given accuracy.

\subsection{Conventions}
We begin with some definitions and conventions.  While we wish to compute the covariance matrix of the configuration space 3PCF, we will end up working in Fourier space to do the computation because simplifications are available there by appeal to the power spectrum.  We define the Fourier transform as
\begin{align}
\tilde{\delta}(\vec{k})=\int d^3 \vec{r} \;\delta(\vec{r})e^{i\vec{k}\cdot\vec{r}}
\end{align}
with inverse 
\begin{align}
\delta(\vec{r})=\int\frac{d^3 \vec{k}}{(2\pi)^3} \;\tilde{\delta}(\vec{k})e^{-i\vec{k}\cdot\vec{r}}.
\end{align}
For the earlier stages of our computation we will in fact need to use the discrete Fourier transform and its inverse, defined as
\begin{align}
\tilde{\delta}(\vec{k})=\sum_r\delta(\vec{r})e^{i\vec{k}\cdot\vec{r}}
\end{align}
and
\begin{align}
\delta(\vec{r})=\frac{1}{V}\sum_k\tilde{\delta}(\vec{k})e^{-i\vec{k}\cdot\vec{r}},
\end{align}
where these discrete transforms are over a volume $V=L^3$ with quantized wavenumbers such that $\vec{k}=2\pi\vec{n}/L$, $n_x,n_y,n_z\in Z$. 

We define the power spectrum as
\begin{align}
P(\vec{k})\delta^K(\vec{k}+\vec{k}')=\frac{1}{V}\left<\tilde{\delta}(\vec{k})\tilde{\delta}(\vec{k}') \right>;
\label{eqn:powerspec_def}
\end{align}
$\delta^K$ is the Kronecker Delta, unity when its argument is zero and zero otherwise. One can check easily that this definition allows one to recover the familiar relation that the correlation function is the Fourier transform of the power spectrum.

We will also use the fact that
\begin{align}
\int d^3 \vec{r} \; e^{i(\vec{k}+\vec{k}')\cdot \vec{r}}=V\delta^K(\vec{k}+\vec{k}').
\end{align}

Finally, note that one can convert from the discrete to the continous case by replacing $(1/V)\sum_k$ with $\int d^3k/(2\pi)^3$.

\subsection{Full covariance}
We now obtain the covariance of our multipole decomposition of the 3PCF.\footnote{ The techniques used here can also be used to compute the covariance of the {\it full} 3PCF, without projection onto the Legendre polynomials, in terms of 2-D integrals, but one obtains an infinite sum over angular momenta. We have not assessed what error truncating this sum might induce.  There may be applications where this 2-D integral representation, despite the infinite sum over angular momenta, could be preferable to the 6-D integral expression of Szapudi (2001).} Here we begin with an estimator for the translation-averaged but not rotation-averaged full 3PCF; we will project onto multipoles (which also averages over rotations) and bin radially later.
\begin{align}
\hat{\zeta}(\vec{r}_1,\vec{r}_2)=\int\frac{d^3\vec{s}}{V} \; \delta(\vec{s})\delta(\vec{s}+\vec{r}_1)\delta(\vec{s}+\vec{r}_2)
\end{align}
For a Gaussian random field, $<\hat{\zeta}>=0$.  The covariance is thus
\begin{align}
&<\hat{\zeta}(\vec{r}_1,\vec{r}_2)\hat{\zeta}(\vec{r}_1',\vec{r}_2')>=\int\frac{d^3\vec{s}d^3\vec{s}'}{V^2}\nonumber\\
&\times\sum_{kqp,k'q'p'}\frac{1}{V^6}\exp\big[-i(\vec{k}\cdot\vec{s}+\vec{q}\cdot(\vec{s}+\vec{r}_1)+\vec{p}\cdot(\vec{s}+\vec{r}_2)\nonumber\\
&+\vec{k}'\cdot\vec{s}'+\vec{q}'\cdot(\vec{s}'+\vec{r}_1')+\vec{p}'\cdot(\vec{s}'+\vec{r}_2')\big]\nonumber\\
&\times \left<\tilde{\delta}(\vec{k})\tilde{\delta}(\vec{q})\tilde{\delta}(\vec{p})\tilde{\delta}(\vec{k}')\tilde{\delta}(\vec{q}')\tilde{\delta}(\vec{p}')\right>.
\end{align}
Peforming the integrals over $d^3\vec{s}$ and $d^3\vec{s}'$ we have
\begin{align}
&<\hat{\zeta}(\vec{r}_1,\vec{r}_2)\hat{\zeta}(\vec{r}_1',\vec{r}_2')>=\sum_{kqp,k'q'p'}\frac{1}{V^6}\delta^K_{kqp}\delta_{k'q'p'}^K\nonumber\\
&\times\exp\big[-i(\vec{q}\cdot\vec{r}_1+\vec{p}\cdot\vec{r}_2+\vec{q}'\cdot\vec{r}_1'+\vec{p}'\cdot\vec{r}_2')\big]\nonumber\\
&\times\left<\tilde{\delta}(\vec{k})\tilde{\delta}(\vec{q})\tilde{\delta}(\vec{p})\tilde{\delta}(\vec{k}')\tilde{\delta}(\vec{q}')\tilde{\delta}(\vec{p}')\right>
\end{align}
where $\delta^K$ is a Kronecker delta whose argument is the sum of the subscripted vectors.
We now use Wick's theorem to reduce the 6-point expectation value to triple products of 2-point functions; this is where Gaussianity enters. We need to consider all possible contractions. 
\begin{align}
&\left<\hat{\zeta}(\vec{r}_{1},\vec{r}_{2})\hat{\zeta}(\vec{r}_{1}',\vec{r}_{2}')\right>\nonumber\\
&=\frac{1}{V^{6}}\sum_{kqp,k'q'p'}\delta_{kqp}^{K}\delta_{k'q'p'}^{K}e^{-i\left[\vec{q}\cdot\vec{r}_{1}+\vec{p}\cdot\vec{r}_{2}+\vec{q}'\cdot\vec{r}_{1}'+\vec{p}'\cdot\vec{r}_{2}'\right]}\nonumber\\
&\times\bigg\{(qq')(pp')(kk')+(pq')(qp')(kk')+(kq')(qk')(pp')\nonumber\\
&+(kp')(qk')(pq')+(kq')(pk')(qp')+(kp')(pk')(qq')\bigg\}
\label{eqn:contracted_covar}
\end{align}
where parentheses represent contractions of $\tilde{\delta}$ evaluated
at the arguments in the parentheses. Using equation (\ref{eqn:powerspec_def}), the term in curly brackets above becomes
\begin{align}
&\bigg\{\cdots\bigg\}=P(q)P(p)P(k)V^{3}\bigg[\delta_{qq'}^{K}\delta_{pp'}^{K}\delta_{kk'}^{K}+\delta_{pq'}^{K}\delta_{qp'}^{K}\delta_{kk'}^{K}\nonumber\\
&+\delta_{kq'}^{K}\delta_{qk'}^{K}\delta_{pp'}^{K}+\delta_{kp'}^{K}\delta_{qk'}^{K}\delta_{pq'}^{K}+\delta_{kq'}^{K}\delta_{pk'}^{K}\delta_{qp'}^{K}+\delta_{kp'}^{K}\delta_{pk'}^{K}\delta_{qq'}^{K}\bigg].
\end{align}
Doing the sums over $k',\;q'$, and $p'$ in equation (\ref{eqn:contracted_covar}) we find 
\begin{align}
&\left<\hat{\zeta}(\vec{r}_{1},\vec{r}_{2})\hat{\zeta}(\vec{r}_{1}',\vec{r}_{2}')\right>=\sum_{kqp}P(q)P(p)P(k)\left(\frac{\delta_{kqp}^{K}}{V^{3}}\right)e^{-i\left[\vec{q}\cdot\vec{r}_{1}+\vec{p}\cdot\vec{r}_{2}\right]}\nonumber\\
&\times\bigg\{ e^{-i\left[\vec{q}\cdot\vec{r}_{1}'+\vec{p}\cdot\vec{r}_{2}'\right]}+e^{-i\left[\vec{p}\cdot\vec{r}_{1}'+\vec{q}\cdot\vec{r}_{2}'\right]}+e^{-i\left[\vec{k}\cdot\vec{r}_{1}'+\vec{p}\cdot\vec{r}_{2}'\right]}\nonumber\\
&+e^{-i\left[\vec{p}\cdot\vec{r}_{1}'+\vec{k}\cdot\vec{r}_{2}'\right]}+e^{-i\left[\vec{k}\cdot\vec{r}_{1}'+\vec{q}\cdot\vec{r}_{2}'\right]}+e^{-i\left[\vec{q}\cdot\vec{r}_{1}'+\vec{k}\cdot\vec{r}_{2}'\right]}\bigg\}.
\label{eqn:sumdone}
\end{align}
Notice each pair of exponentials in the curly brackets is obviously symmetric under switching $\vec{r}_{1}'\leftrightarrow\vec{r}_{2}'$.
Also notice from the first line that equation (\ref{eqn:sumdone}) is symmetric under
$\vec{r}_{1}\leftrightarrow\vec{r}_{2}$ if we also flip $\vec{q}$
and $\vec{p}$. Applying this to all of the terms in curly brackets
too, we find ${b+a+e+f+c+d}$ if we had originally labeled each exponential in the curly brackets as ${a+b+c+d+e+f}.$
Hence the equation has the desired symmetries.

Converting this into an integral we have
\begin{align}
&{\rm Cov}\equiv\left<\hat{\zeta}(\vec{r}_{1},\vec{r}_{2})\hat{\zeta}(\vec{r}_{1}',\vec{r}_{2}')\right>=\frac{1}{V}\int\frac{d^3\vec{q}d^3\vec{p}d^3\vec{k}}{\left(2\pi\right)^{9}}\;P(p)P(q)P(k)\nonumber\\
&\times \left(2\pi\right)^{3}\delta_{D}^{[3]}\left(\vec{q}+\vec{p}+\vec{k}\right)e^{-i\left[\vec{q}\cdot\vec{r}_{1}+\vec{p}\cdot\vec{r}_{2}\right]}\bigg\{ \cdots\bigg\},
\label{eqn:fullcovar}
\end{align}
where above we have not rewritten the terms in curly brackets from equation (\ref{eqn:sumdone}).

\subsection{Projection onto Legendre polynomials}
\label{subsec:projection}
We now consider the covariance projected onto multipoles, defining
\begin{align}
&{\rm Cov}_{ll'}(r_1,r_2;r_1',r_2')=\frac{(2l+1)(2l'+1)}{(4\pi)^4}\int d\Omega_{r1}d\Omega_{r2}d\Omega_{r1'}d\Omega_{r2'}\nonumber\\
&\times {\rm Cov}(\vec{r}_1,\vec{r}_2;\vec{r}_1',\vec{r}_2')P_l(\hat{r}_1\cdot\hat{r}_2)P_{l'}(\hat{r}_1'\cdot\hat{r}_2').
\label{eqn:proj_covar_def}
\end{align}
Noticing that in equation (\ref{eqn:fullcovar}) the exponentials contain the only $\vec{r}$ dependence, we first define the projection of one exponential onto one Legendre polynomial as
\begin{align}
&I_{{\rm proj},l}(\vec{r}_{1},\vec{k}_{1};\vec{r}_{2},\vec{k}_{2})\nonumber\\
&=\frac{2l+1}{\left(4\pi\right)^{2}}\int e^{-i\left[\vec{k}_{1}\cdot\vec{r}_{1}+\vec{k}_{2}\cdot\vec{r}_{2}\right]}P_{l}(\hat{r}_{1}\cdot\hat{r}_{2})d\Omega_{r1}d\Omega_{r2}\nonumber\\
&=(2l+1)(-1)^{l}\mathcal{J}_{l}(k_{1},k_{2})P_{l}(\hat{k}_{1}\cdot\hat{k}_{2}).
\label{eqn:Iproj_def}
\end{align}
We will have this factor from projecting the exponential outside the curly brackets in equation (\ref{eqn:fullcovar}),
and then six analogous factors within the curly brackets from projecting each exponential of $\vec{r}_1',\vec{r}_2'$ onto $P_{l'}(\hat{r}_1'\cdot\hat{r}_2')$.

We have defined $\mathcal{J}_{l}(x,y)=j_{l}(xr_{1})j_{l}(yr_{2})$
and will also use $\mathcal{J}'_{l'}(x,y)=j_{l'}(xr_{1}')j_{l'}(yr_{2}')$.
We performed the projection integral by expanding the exponential in spherical harmonics
using AWH13 equation (16.61) and expanding the Legendre
polynomial in spherical harmonics using the spherical harmonic addition
theorem (\ref{eqn:sph_addition_theorem}); the integral can then be evaluated by orthogonality.

\begin{figure}
\includegraphics[scale=0.35]{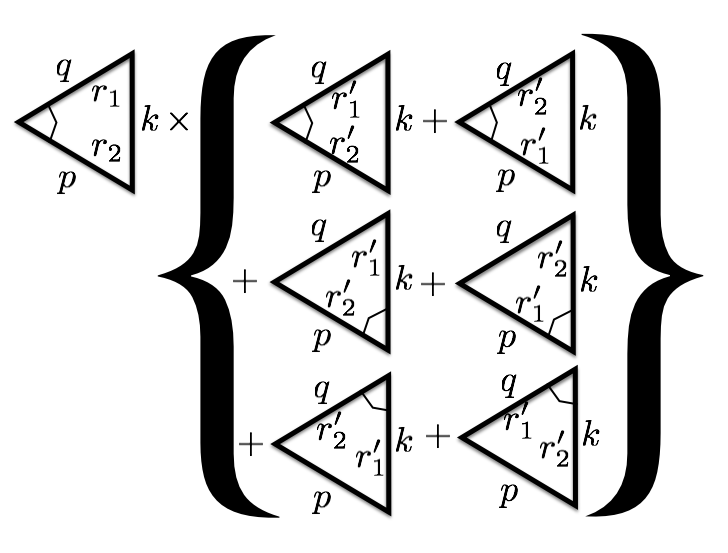}
\caption{This shows the symmetries of equation (\ref{eqn:proj_covar_qp}), which in turn derive from the structure of equation (\ref{eqn:sumdone}); one can directly compare the arguments of the exponentials in this latter with the diagram.  The leftmost triangle represents the term $\mathcal{J}_l(q,p)P_l(\hat{q}\cdot\hat{p})$ outside the curly brackets in equation (\ref{eqn:proj_covar_qp}), showing also the radial arguments implicit in the $\mathcal{J}_l$, and the six triangles inside the curly brackets above represent the six terms in the curly brackets.  The Legendre polynomials are always evaluated about a particular vertex, as shown in the diagram, and the real-space variables match to Fourier-space variables differently in each triangle (see equation (\ref{eqn:sumdone})).  One can see from above that each pair of triangles, or pair of terms in curly brackets in equation (\ref{eqn:proj_covar_qp}), has switch symmetry $\vec{r}_1'\leftrightarrow\vec{r}_2'$.  These are rotation symmetries about the vertex between $\vec{r}_1'$ and $\vec{r}_2'$ in each pair.  One can also see that if we switch $\vec{r}_1\leftrightarrow\vec{r}_2$ and $\vec{q}\leftrightarrow \vec{p}$, the leftmost triangle is symmetric.  The topmost pair will also be symmetric under this switch as well, which is why it gives rise to two symmetric projection integrals, but the middle and bottom pairs will not be, which is why they give rise to four asymmetric projection integrals (see equation (\ref{eqn:covar_ito_ints})).}
\label{fig:covar_cycles_diagram}
\end{figure}

Writing out the projection integrals explicitly using equation (\ref{eqn:Iproj_def}), we thus have the projected covariance as
\begin{align}
&{\rm Cov}_{ll'}=\frac{1}{V}\int\frac{d^3\vec{q}d^3\vec{p}d^3\vec{k}}{\left(2\pi\right)^{9}} \; P(p)P(q)P(k)\left(2\pi\right)^{3}\delta_{D}^{[3]}(\vec{p}+\vec{q}+\vec{k})\nonumber\\
&\times(2l+1)(2l'+1)(-1)^{l+l'}\mathcal{J}_{l}(q,p)P_{l}(\hat{q}\cdot\hat{p})\nonumber\\
&\times\bigg\{\mathcal{J}'_{l'}(q,p)P_{l'}(\hat{q}\cdot\hat{p})+\mathcal{J}'_{l'}(p,q)P_{l'}(\hat{p}\cdot\hat{q})+\mathcal{J}'_{l'}(k,p)P_{l'}(\hat{k}\cdot\hat{p})\nonumber\\
&+\mathcal{J}'_{l'}(p,k)P_{l'}(\hat{p}\cdot\hat{k})+\mathcal{J}'_{l'}(k,q)P_{l'}(\hat{k}\cdot\hat{q})+\mathcal{J}'_{l'}(q,k)P_{l'}(\hat{q}\cdot\hat{k})\bigg\}.
\label{eqn:proj_covar_qp}
\end{align}
Note that in equation (\ref{eqn:proj_covar_qp}) the Legendre polynomial dependence is the same for each of the first
pair in the curly brackets, the second pair, and the third pair because
the dot product is symmetric. Thus we have three possible angular integrals
to do, corresponding to these three pairs:
\begin{align}
&I_{{\rm ang},ll'}^{{\rm symm}}=\int d\Omega_{p}d\Omega_{q}d\Omega_{k}P_{l}(\hat{q}\cdot\hat{p})P_{l'}(\hat{q}\cdot\hat{p})(2\pi)^{3}\delta_{D}^{[3]}(\vec{k}+\vec{p}+\vec{q})\nonumber\\
&I_{{\rm ang},ll'}^{{\rm asymm}}=\int d\Omega_{p}d\Omega_{q}d\Omega_{k}P_{l}(\hat{q}\cdot\hat{p})P_{l'}(\hat{k}\cdot\hat{p})(2\pi)^{3}\delta_{D}^{[3]}(\vec{k}+\vec{p}+\vec{q})\nonumber\\
&I_{{\rm ang},ll'}^{{\rm asymm}}=\int d\Omega_{p}d\Omega_{q}d\Omega_{k}P_{l}(\hat{q}\cdot\hat{p})P_{l'}(\hat{k}\cdot\hat{q})(2\pi)^{3}\delta_{D}^{[3]}(\vec{k}+\vec{p}+\vec{q}).
\label{eqn:projxn_integrals}
\end{align}
Note that the second and third integrals above are really the same
under $\vec{p}\leftrightarrow\vec{q}$. We term the first integral
above the symmetric integral and the second and third asymmetric. Figure \ref{fig:covar_cycles_diagram} explains these equations and their symmetries diagrammatically to illustrate the underlying structure of the covariance calculation up to this point.

To evaluate these angular integrals, we write the Dirac delta as the Fourier transform of unity,
\begin{equation}
\left(2\pi\right)^{3}\delta_{D}^{[3]}(\vec{k}+\vec{p}+\vec{q})=\int d^3\vec{r}\;e^{i\left[\vec{k}\cdot\vec{r}+\vec{p}\cdot\vec{r}+\vec{q}\cdot\vec{r}\right]}
\end{equation}
expand each exponential in spherical harmonics using AWH13 equation (16.61), and perform the angular integral
over $d\Omega_{r}$. Defining
\begin{align}
&\mathcal{R}_{l_{1}l_{2}l_{3}}(k,p,q)=\int r^{2}drj_{l_{1}}(kr)j_{l_{2}}(pr)j_{l_{3}}(qr),\nonumber\\
&\mathcal{D}_{l_{1}l_{2}l_{3}}=i^{l_{1}+l_{2}+l_{3}}\\
&\mathcal{C}_{l_{1}l_{2}l_{3}}=\sqrt{\frac{(2l_{1}+1)(2l_{2}+1)(2l_{3}+1)}{4\pi}},\nonumber
\end{align}
we obtain
\begin{align}
&\left(2\pi\right)^{3}\delta_{D}^{[3]}(\vec{k}+\vec{p}+\vec{q})=\nonumber\\
&\left(4\pi\right)^{3}\sum_{l_{1}l_{2}l_{3},m_{1}m_{2}m_{3}}\mathcal{D}_{l_{1}l_{2}l_{3}}\mathcal{C}_{l_{1}l_{2}l_{3}}\mathcal{R}_{l_{1}l_{2}l_{3}}(k,p,q)\nonumber\\
&\times\left(\begin{array}{ccc}
l_{1} & l_{2} & l_{3}\\
0 & 0 & 0
\end{array}\right)\left(\begin{array}{ccc}
l_{1} & l_{2} & l_{3}\\
m_{1} & m_{2} & m_{3}
\end{array}\right)\nonumber\\
&\times Y_{l_{1}m_{1}}^{*}(\hat{k})Y_{l_{2}m_{2}}^{*}(\hat{p})Y_{l_{3}m_{3}}^{*}(\hat{q}).
\label{eqn:delta_as_ylms}
\end{align}
This is equivalent to Mehrem (2002) equation (5.1) if the 3j-symbols above are translated to Clebsch-Gordan symbols.

Inserting equation (\ref{eqn:delta_as_ylms}) into equation (\ref{eqn:projxn_integrals}) and then expanding the Legendre polynomials in equation (\ref{eqn:projxn_integrals}) into spherical harmonics using the spherical harmonic addition theorem (\ref{eqn:sph_addition_theorem}), we now simply have integrals over products of three spherical harmonics, which can be done analytically with 3j-symbols. The result can then be simplified by explicitly evaluating some of the 3j-symbols (using NIST Digital Library of Mathematical Functions (DLMF) 34.3.1) and summing over all of the spin angular momenta (using NIST DLMF 34.3.10 and 34.3.18). For the symmetric integral we find
\begin{align}
&I_{{\rm ang},ll'}^{{\rm symm}}(p,q;k)=\nonumber\\
&(4\pi)^{4}\sum_{l_{2}}(-1)^{l_2}(2l_{2}+1)\left(\begin{array}{ccc}
l & l' & l_{2}\\
0 & 0 & 0
\end{array}\right)^{2}\mathcal{R}_{l_{2}l_{2}0}(p,q,k)
\label{eqn:symm_int}
\end{align}
where we have separated $k$ with a semicolon because it is the only
argument that does not appear in the Legendre polynomials in the integral. A simple case to check is setting $l'=0$ and $k=0$ in equation (\ref{eqn:projxn_integrals}). Then $\vec{q} = -\vec{p}$ so by direct computation
\begin{align}
\lim_{k\to 0}I^{\rm symm}_{{\rm ang},l0}=128\pi^5(-1)^l \frac{\delta_{\rm D}^{[1]}(p-q)}{q^2}
\label{eqn:symm_direct}
\end{align}
 where we used $P_l(-1)=(-1)^l$. In equation (\ref{eqn:symm_int}), $l'=0$ sets $l_2 =l$ and the 3j-symbol's square is $1/(2l+1)$.  Using the orthogonality relation for spherical Bessel functions, $\mathcal{R}_{l l 0}(p,q,0)=\pi \delta_{\rm D}^{[1]}(p-q)/(2q^2)$; inserting this in equation (\ref{eqn:symm_int}) and simplifying yields agreement with the direct computation.

For the asymmetric integral we find
\begin{align}
&I_{{\rm ang},ll'}^{{\rm asymm}}(p;q,k)=\left(4\pi\right)^{4}\nonumber\\
&\times\sum_{l_{2}}(2l_{2}+1)\left(\begin{array}{ccc}
l & l' & l_{2}\\
0 & 0 & 0
\end{array}\right)^{2}(-1)^{(l+l'+l_{2})/2}\mathcal{\mathcal{R}}_{l_{2}ll'}(p,q,k),
\label{eqn:asymm_int}
\end{align}
where now $p$ is separated by a semicolon because it appeared in two Legendre polynomials in the integrand. Note that for $l'=0$, the symmetric and asymmetric integrals of equation (\ref{eqn:projxn_integrals}) are equal, so equation (\ref{eqn:asymm_int}) should reduce to equation (\ref{eqn:symm_int}) in this limit, as can be verified by noting $l'=0$ implies $l=l_2$.

Thus
\begin{align}
&{\rm Cov}_{ll'}=\frac{1}{V}\int\frac{q^{2}p^{2}k^{2}}{(2\pi)^{9}}P(p)P(q)P(k)(2l+1)(2l'+1)(-1)^{l+l'}\nonumber\\
&\times\mathcal{J}_{l}(q,p)\bigg\{\mathcal{J}'_{l'}(q,p)I_{{\rm ang},ll'}^{{\rm symm}}(q,p;k)+\mathcal{J}'_{l'}(p,q)I_{{\rm ang},ll'}^{{\rm symm}}(p,q;k)\nonumber\\
&+\mathcal{J}'_{l'}(k,p)I_{{\rm ang},ll'}^{{\rm asymm}}(p;q,k)+\mathcal{J}'_{l'}(p,k)I_{{\rm ang},ll'}^{{\rm asymm}}(p;q,k)\nonumber\\
&+\mathcal{J}'_{l'}(k,q)I_{{\rm ang},ll'}^{{\rm asymm}}(q;p,k)+\mathcal{J}'_{l'}(q,k)I_{{\rm ang},ll'}^{{\rm asymm}}(q;p,k)\bigg\}.
\label{eqn:covar_ito_ints}
\end{align}

We now interchange the order of integration so that the integrals
over $q,p$ and $k$ are done first, since they are separable, and
the linking integral over $r$ implied by $\mathcal{R}_{l_{1}l_{2}l_{3}}$
is done last. We also make the sum over $l_{2}$ explicit and do it
after evaluating the $q,p$ and $k$ integrals. Finally we define 
\begin{equation}
f_{ll}(r;r_{1})=\int\frac{k^{2}dk}{2\pi^{2}}P(k)j_{l}(kr_{1})j_{l}(kr)
\end{equation}
and
\begin{equation}
f_{l_{2}ll'}(r;r_{1},r_{1}')=\int\frac{k^{2}dk}{2\pi^{2}}P(k)j_{l}(kr_{1})j_{l'}(kr_{1}')j_{l_{2}}(kr).
\end{equation}
In terms of these functions,
\begin{align}
&{\rm Cov}_{ll'}(r_{1},r_{2};r_{1}',r_{2}')=\frac{4\pi}{V}(2l+1)(2l'+1)(-1)^{l+l'}\nonumber\\
&\times\int r^{2}dr\sum_{l_{2}}(2l_{2}+1)\left(\begin{array}{ccc}
l & l' & l_{2}\\
0 & 0 & 0
\end{array}\right)^2\nonumber\\
&\times\bigg\{(-1)^{l_2}\xi_{0}(r)\bigg[f_{l_{2}ll'}(r;r_{1},r_{1}')f_{l_{2}ll'}(r;r_{2},r_{2}')\nonumber\\
&+f_{l_{2}ll'}(r;r_{2},r_{1}')f_{l_{2}ll'}(r;r_{1},r_{2}')\bigg]+(-1)^{(l+l'+l_{2})/2}\nonumber\\
&\times\bigg[f_{ll}(r;r_{1})f_{l'l'}(r;r_{1}')f_{l_{2}ll'}(r;r_{2},r_{2}')\nonumber\\
&+f_{ll}(r;r_{1})f_{l'l'}(r;r_{2}')f_{l_{2}ll'}(r;r_{2},r_{1}')\nonumber\\
&+f_{ll}(r;r_{2})f_{l'l'}(r;r_{1}')f_{l_{2}ll'}(r;r_{1},r_{2}')\nonumber\\
&+f_{ll}(r;r_{2})f_{l'l'}(r;r_{2}')f_{l_{2}ll'}(r;r_{1},r_{1}')\bigg]\bigg\}.
\label{eqn:fullcovar_final}
\end{align}
One can see that this is symmetric under switching $r_1\leftrightarrow r_2$ and $r'_1\leftrightarrow r'_2$, as expected.  We have thus shown how to reduce the covariance of our multipole decomposition to a sum of 2-D integrals. This is a significant computational benefit: the $f_{ll}$ and $f_{l_2 l l'}$ can be pre-computed once to give all the terms in the sum above, and then integrated over $dr$. Further, since the problem is now 2-D one can simply evaluate the integrals using a grid and avoid appealing to more complicated higher-dimensional integration techniques.

In closing, we note that for $P\propto 1/k$ (i.e. $\xi_0\propto 1/r^2$), $f_{ll}$ and $f_{l_2ll'}$ can be computed analytically. We find
\begin{align}
f_{ll}(r;r_1)=\frac{1}{4\pi^{3/2} }r^l r_1^{-l-2}\frac{\Gamma(l+1)}{\Gamma(l+\frac{3}{2})}\nonumber\\
\times  F\bigg(l+1,\frac{1}{2};l+\frac{3}{2};\left(\frac{r}{r_1}\right)^2\bigg)
\end{align}
using $j_l(x)=\sqrt{\pi/(2x)}J_{l+1/2}(x)$ and Gradshteyn \& Ryzhik (2007) equation (6.512.1). $F$ is the hypergeometric function and we assume $r<r_1$; the result for $r>r_1$ is given by switching $r_1 \leftrightarrow r$.
$f_{l_2ll'}$ can be computed using techniques outlined in Fabrikant (2013) and is given by his equation (9); since the expression is rather long we do not reproduce it here. We mention this since one could imagine scenarios in which high speed was desirable for computing the covariance, such that these approximate forms might suffice. Finally, to incorporate shot noise in the covariance, one takes $P(k)\to P(k)+1/n$, $n$ the survey number density.  This will introduce a cross term where one of the $f_{ll}$ or $f_{l_2ll'}$ in each pair in equation (\ref{eqn:fullcovar_final}) no longer involves the power spectrum, and also a term in $1/n^2$ where both functions in each pair do not. The required integrals are also analytic:
\begin{align}
f_{ll}(r;r_1)\to  \frac{1}{4\pi n r_1^2}\delta_{\rm D}^{[1]}(r-r_1),
\end{align}
while $f_{l_2ll'}$ is rather longer and given by Mehrem (2002) equation (5.14), assuming a much simpler form (his equation (5.15)) if $l,l'$, or $l_2=0$.

\subsection{Radial binning for the covariance}
\label{subsec:binning_covar}
The above calculation used exact values for $r_1,\;r_2,\;r_1'$ and $r_2'$, but we can easily integrate over bins in radius. We simply integrate the $f_{l_2ll'}$ and $f_{ll}$ functions defined above over bins, equivalent to replacing $j_l(kr_1)$ and $j_l(kr_1')$ with their bin-averaged values.  We define
\begin{equation}
\bar{f}_{ll}(r_i)=\int\frac{k^{2}dk}{2\pi^{2}}P(k)\bar{j}_{l}(k;r_i)j_{l}(kr)
\label{eqn:flbar_2}
\end{equation}
and
\begin{equation}
\bar{f}_{l_{2}ll'}(r;r_i,r'_i)=\int\frac{k^{2}dk}{2\pi^{2}}P(k)\bar{j}_{l}(k;r_i)\bar{j}_{l'}(k;r'_i)j_{l_{2}}(kr)
\label{eqn:flbar_3}
\end{equation}
with 
\begin{equation}
\bar{j}_l(k;r_i)=\frac{\int u^2du \; j_l(ku)\Phi(u;r_i)}{\int u^2 du \;\Phi(u;r_i)}
\label{eqn:jlbar}
\end{equation}
where $u$ is a dummy variable and we recall that $\Phi(u;r_i)$ is the binning function ensuring $u$ is in the bin $r_i$ (see Section \ref{subsec:binning}).

\subsection{Covariance results}
\label{subsec:covresults}
\begin{figure*}
\centering
\includegraphics[scale=.8]{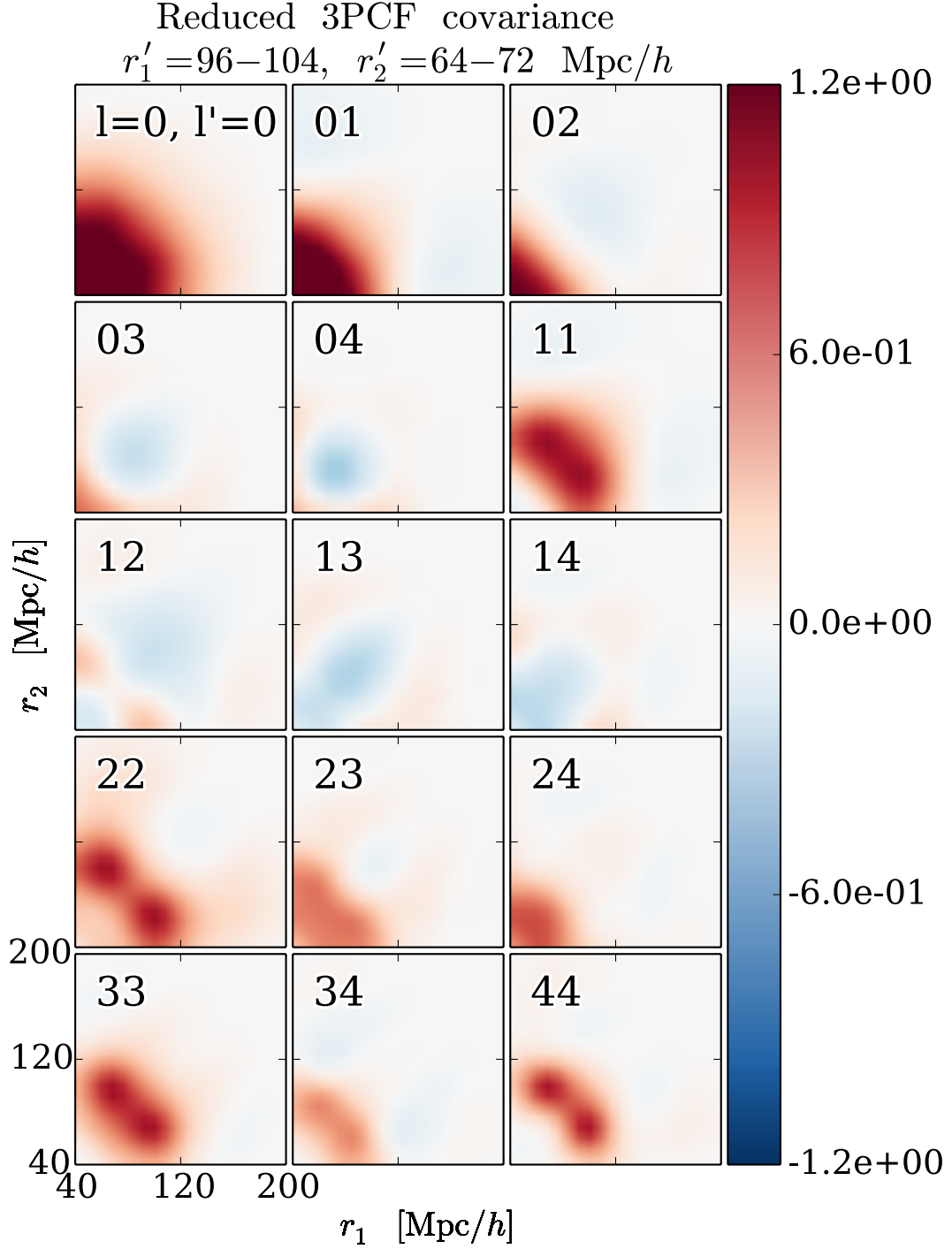}
\caption{The reduced covariance (equation (\ref{eqn:red_covar})) for a number of multipole combinations.  The $r_1'$ and $r_2'$ bins are fixed, and the $r_1$ and $r_2$ bins are the horizontal and vertical axes of the plot; the $ll'$ combination is indicated in the upper left of each panel. We have chosen an $(r_1', r_2')$ bin combination for relevance to the BAO scale ($\sim 100{\rm\; Mpc}/h$) while avoiding the squeezed limit (hence $r_2'$ away from $r_1'$).}
\label{fig:covar_grid}
\end{figure*}

We display the binned reduced covariance,
\begin{align}
&{\rm Red\;Cov}_{ll'}(r_1,r_2;r_1',r_2')=\nonumber\\
&\frac{{\rm Cov}_{ll'}(r_1,r_2;r_1',r_2')}{\sqrt{{\rm Cov}_{ll}(r_1',r_2';r_1',r_2'){\rm Cov}_{l'l'}(r_1',r_2';r_1',r_2')}}
\label{eqn:red_covar}
\end{align}
for fixed $r_1',r_2'$ and a number of $r_1,r_2$ and $ll'$ combinations in Figure \ref{fig:covar_grid}.  One can see clear features when $(r_1,r_2)=(r_1',r_2')$, especially when $l=l'$ as well (e.g. the 11, 22, 33, and 44 panels). The computation was done in $8{\rm\; Mpc}/h$ bins but we display with an interpolated color scheme because the underlying radial variables are continuous, in contrast to the multipoles $l$ and $l'$.  We used the linear-theory matter power spectrum from CAMB (Lewis 2000) and checked convergence of the integrals by varying the endpoints and spacing of the grids in $r$ and $k$ we used.\footnote{Here we used the redshift zero power spectrum from a flat $\Lambda CDM$ cosmology with $\Omega_{\rm b}h^2=0.0226,\;\Omega_{\rm c}h^2=0.112,\;n_s=0.96$, and $\sigma_8=0.821$.} For the spherical Bessel functions we used high-order Taylor series for small values of the arguments, with the change-over point to the series depending on the order $l$, and cross-checked with direct computation using {\it scipy's} built-in functions. For the  $\bar{j}_l$ (equation (\ref{eqn:jlbar})) we used analytical results, cross-checked with numerical integrations of the $j_l$.

\begin{figure*}
\centering
\includegraphics[scale=.8]{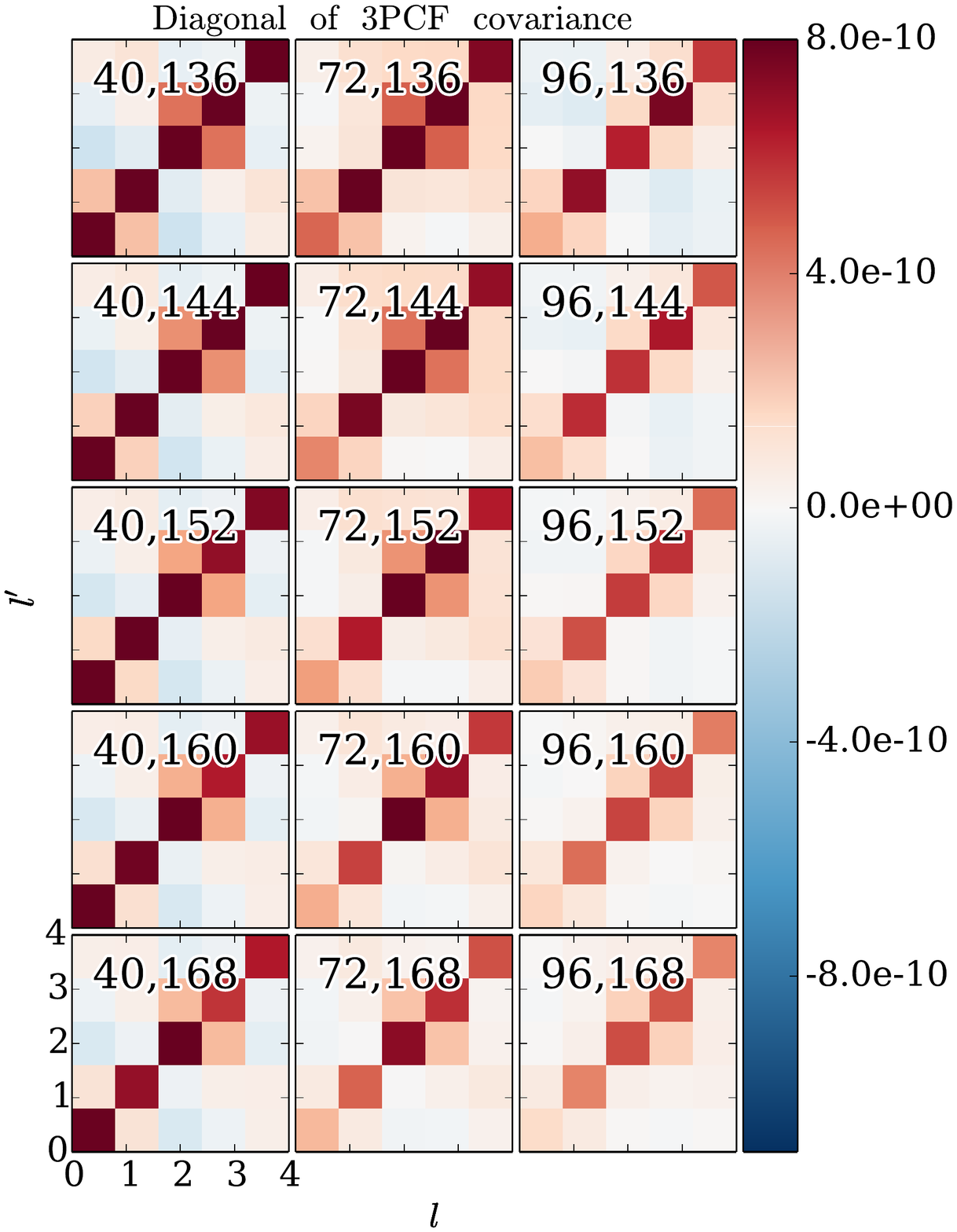}
\caption{The diagonal of the binned covariance, i.e. equation (\ref{eqn:fullcovar_final}) binned and with $(r_1,r_2)=(r_1',r_2')$. The $(r_1,r_2)$ bins are indicated in the upper left of each panel, and the horizontal and vertical axes show the multipoles $l$ and $l'$. We have used a survey volume $V$ in equation (\ref{eqn:fullcovar_final}) of $1\;({\rm Gpc}/h)^3$, roughly that of the SDSS DR12, to normalize. This is about 7 times smaller than the total volume of mock catalogs used here; thus $1/\sqrt{7}$ times the square-root of the diagonals above gives a rough estimate of the error bars we might expect as of order $10^{-5}$, or about $5\%$ of the 3PCF (comparing with Figure \ref{fig:3PCF_colorgrid}).}
\label{fig:covar_diag_grid}
\end{figure*}

In Figure \ref{fig:covar_diag_grid}, we show the binned covariance when $r_1=r_1'$ and $r_2=r_2'$ versus all $l$ and $l'$, and for a number of choices of $r_1'$ and $r_2'$, indicated in the upper left of each panel. Notice that the strongest covariance is, as one might expect, when $l=l'$ as well, along the diagonal.  We display with no color interpolation because the multipoles are discrete. We show larger radial bins than the LasDamas mock results contain because these will be relevant for the Baryon Acoustic Oscillation (BAO) scale analysis planned for future work. 

\section{Mock data results}
\label{sec:mock_results}
\subsection{Full results}
\label{subsec:full_results}
\begin{figure*}
\centering
\includegraphics[scale=.8]{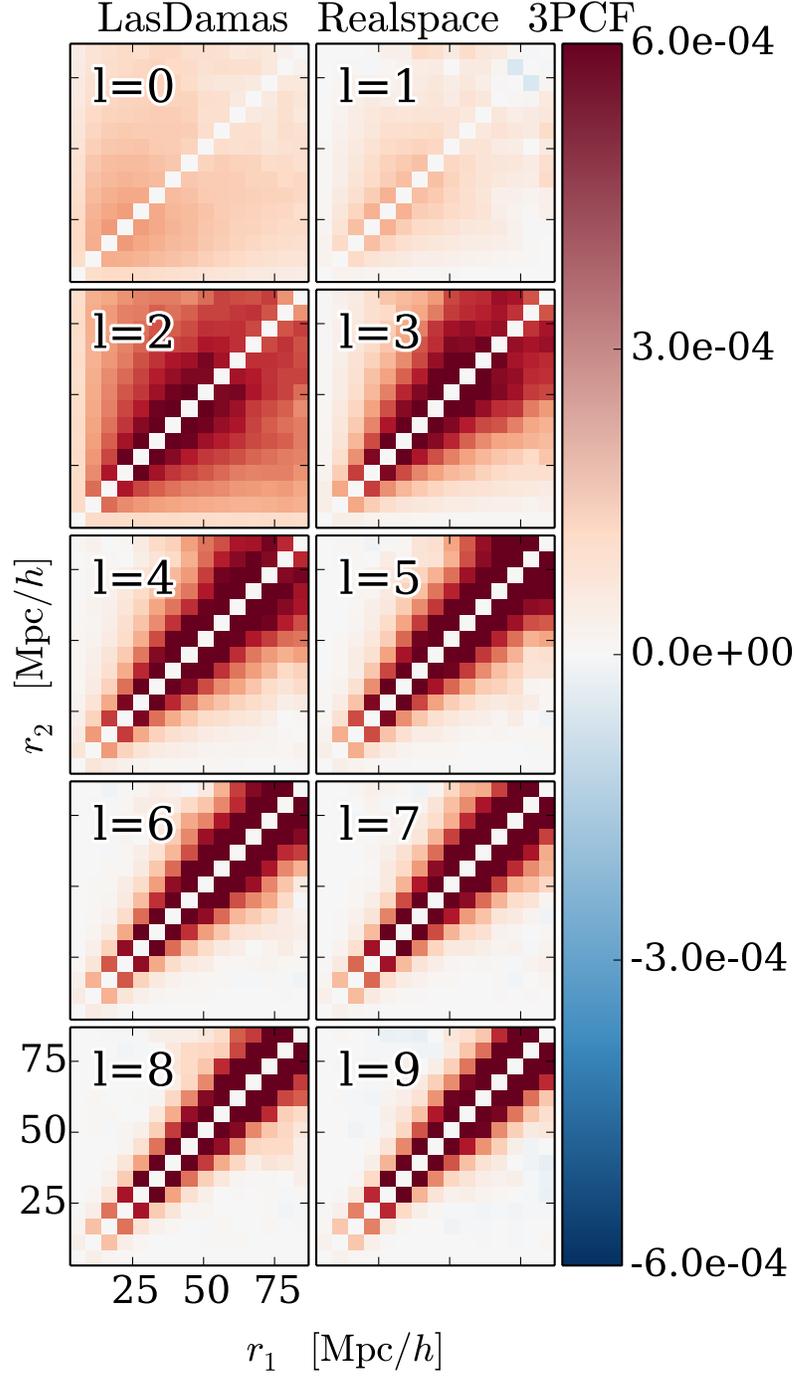}
\caption{Binned multipole coefficients $\bar{\zeta}_l$ of the 3PCF as defined in equations (\ref{eqn:zeta_series}) and (\ref{eqn:zeta_binned_final}).  These are the result of applying our algorithm to the LasDamas real space mock catalogs for the SDSS DR7, as described in Section \ref{sec:mock_results}, and with edge correction as described in Section \ref{sec:edgecorrxn}. The horizontal and vertical axes are $r_1$ and $r_2$ in ${\rm Mpc}/h$. As in SE15, we have weighted by the volume in each spherical shell $r_1^2r_2^2/(100\;{\rm Mpc}/h)^4$; this is to amplify the finer features. Note that on the diagonal, we expect the 3PCF to be dominated by squeezed triangles for which perturbation theory is not valid, so we have not computed the diagonal. For this reason we need not include the additional exponential suppression of the diagonal used in SE15.}
\label{fig:3PCF_colorgrid}
\end{figure*}

We present the results of running our algorithm on the publicly available LasDamas mock catalogs for the SDSS-II DR7 in both real and redshift space.\footnote{\url{http://lss.phy.vanderbilt.edu/lasdamas/mocks/}} We used 15 radial bins with $\Delta r =6\; {\rm Mpc}/h$.  We show first the results at each multipole versus the two triangle side lengths $r_1$ and $r_2$ in Figure \ref{fig:3PCF_colorgrid}. This shows that the largest amplitude contribution to the 3PCF, especially for triangles well away from the diagonal, is $l=2$. This is what we expect from SE15, Figure 9, third row, leftmost panel, showing the perturbation theory results and focusing on the linear bias $b_1$, which dominates the non-linear bias $b_2$. In $l=1$, there is a hint of a large-scale decrement, to be compared with the slight feature close to the diagonal around $130\;{\rm Mpc}(=90\;{\rm Mpc}/h)$ in SE15 Figure 9, second row, leftmost panel. 

As in SE15, $l=2$ and $l=3$ look similar but with $l=2$ having higher amplitude away from the diagonal. For $l\geq 3$, the panels all begin to look the same, agreeing with our expectation from SE15 Figure 9.  This is because these higher multipoles, in particular near the diagonal, are dominated by a small population of squeezed triangles where two sides are equal (e.g. $r_1$ and $r_2$) and the third side nears zero.  In the hierarchical ansatz for the 3PCF, one has $\zeta\sim 1/(r_1r_2)^2+1/(r_2r_3)^2+1/(r_3 r_1)^2$, so we expect the amplitude to become very large as any side approaches zero. 

Also discussed in SE15 is another reason for the similarity of the $l\geq 3$ panels: before cyclic summing over vertices of the triangle, the  leading order pre-cyclic perturbation theory 3PCF only has structure for $l=0,1,2$.  In particular, at leading (fourth) order, the 3PCF receives one contribution from the second-order density field, $\delta^{(2)}$, which is in turn calculated by integrating a kernel $\tilde{F}^{(2)}$ against the linear density field (Goroff et al. 1986; Jain \& Bertschinger 1994; Bernardeau et al. 2002).  This kernel has only $l=0,1$, and $2$ terms.  If one chooses the second-order density point to be at the origin, the 3PCF therefore has multipole structure only to $l=2$.  In reality we do not know which point contributes $\delta^{(2)}$, so we must cyclically sum around the triangle and cannot choose $\delta^{(2)}$ at the origin.  This cyclic summing generates additional angular structure, but it just stems from the geometric effect of writing a simple $l=0,1,$ and $2$-only multipole expansion with argument e.g. $\hat{r}_1\cdot\hat{r}_3=\cos\theta_{13}$ in terms of $\hat{r}_1\cdot\hat{r}_2=\cos\theta_{12}$. 

We note that the $l=0,1$, and $2$ terms that enter pre-cyclically have a physical meaning. $\tilde{F}^{(2)}$ is formed by summing two mode-coupling kernels $\alpha$ and $\beta$ (Bernardeau et al. 2002 equations (39) and (156)). These in turn come from solving respectively the full continuity equation and the Euler equation (compare Bernardeau et al. 2002 equations (16) and (17) with their equations (37) and (38)).  $\alpha$ produces all of the $l=0$ and $5/7$ of the $l=1$ terms in $\tilde{F}^{(2)}$. The $l=0$ contribution is from the product of the velocity divergence and the density, while the $l=1$ contribution is from gradients of the density field parallel to the velocity.  Meanwhile, $\beta$ generates the remaining $2/7$ of the $l=1$ term and all of the $l=2$ term in $\tilde{F}^{(2)}$; these stem from gradients of the velocity divergence parallel to the velocity.

\begin{figure}
\includegraphics[scale=0.4]{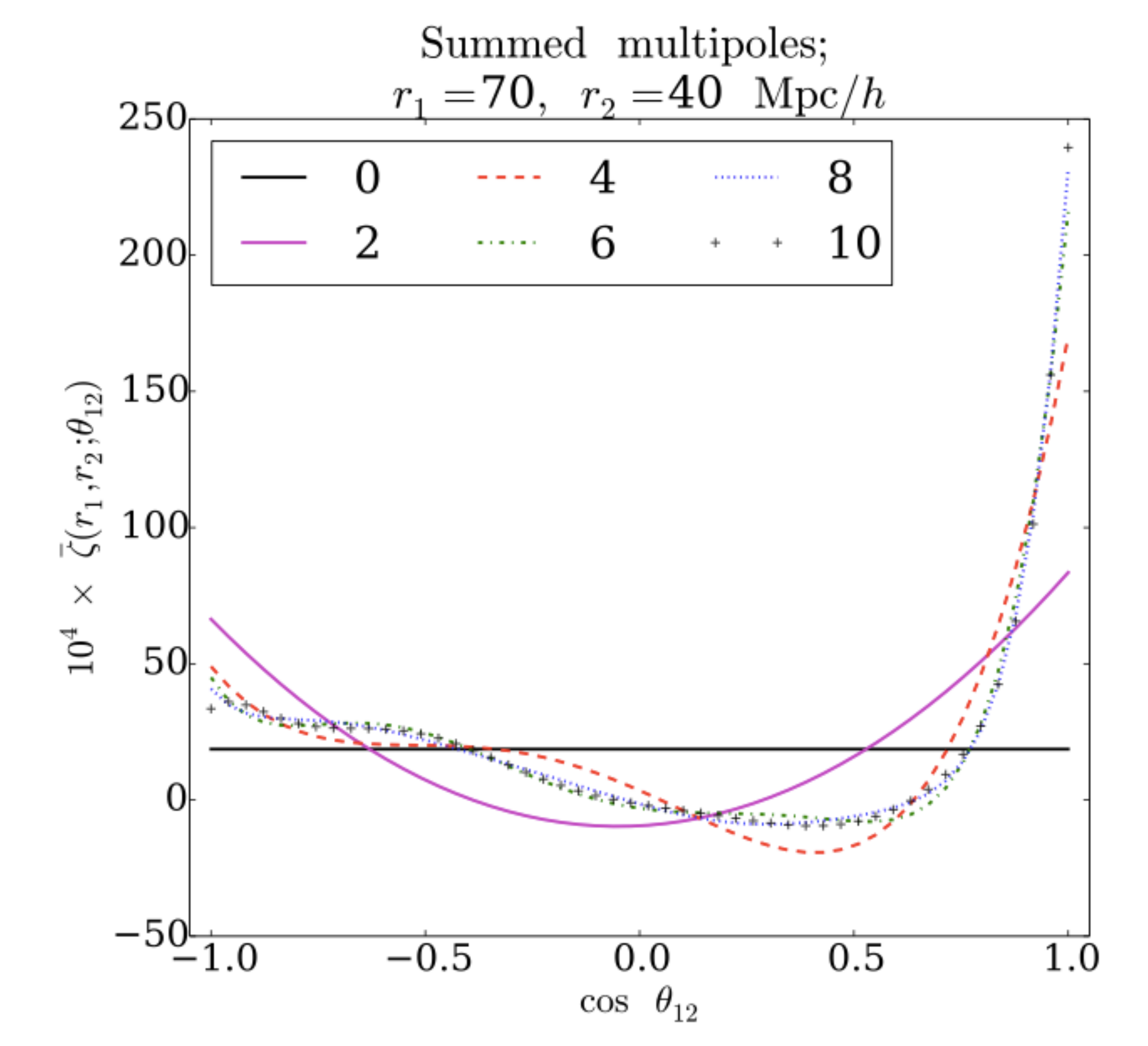}
\caption{This shows reconstruction of the full 3PCF (for the LasDamas real space mocks) from its multipoles using coefficients up to and including the $l$ indicated in the legend, as described in Section \ref{sec:mock_results}.  The reconstruction converges even for modest $l$. The $l=8$ points lie essentially directly under those for $l=10$.}
\label{fig:recon}
\end{figure}

\begin{figure}
\includegraphics[scale=0.35]{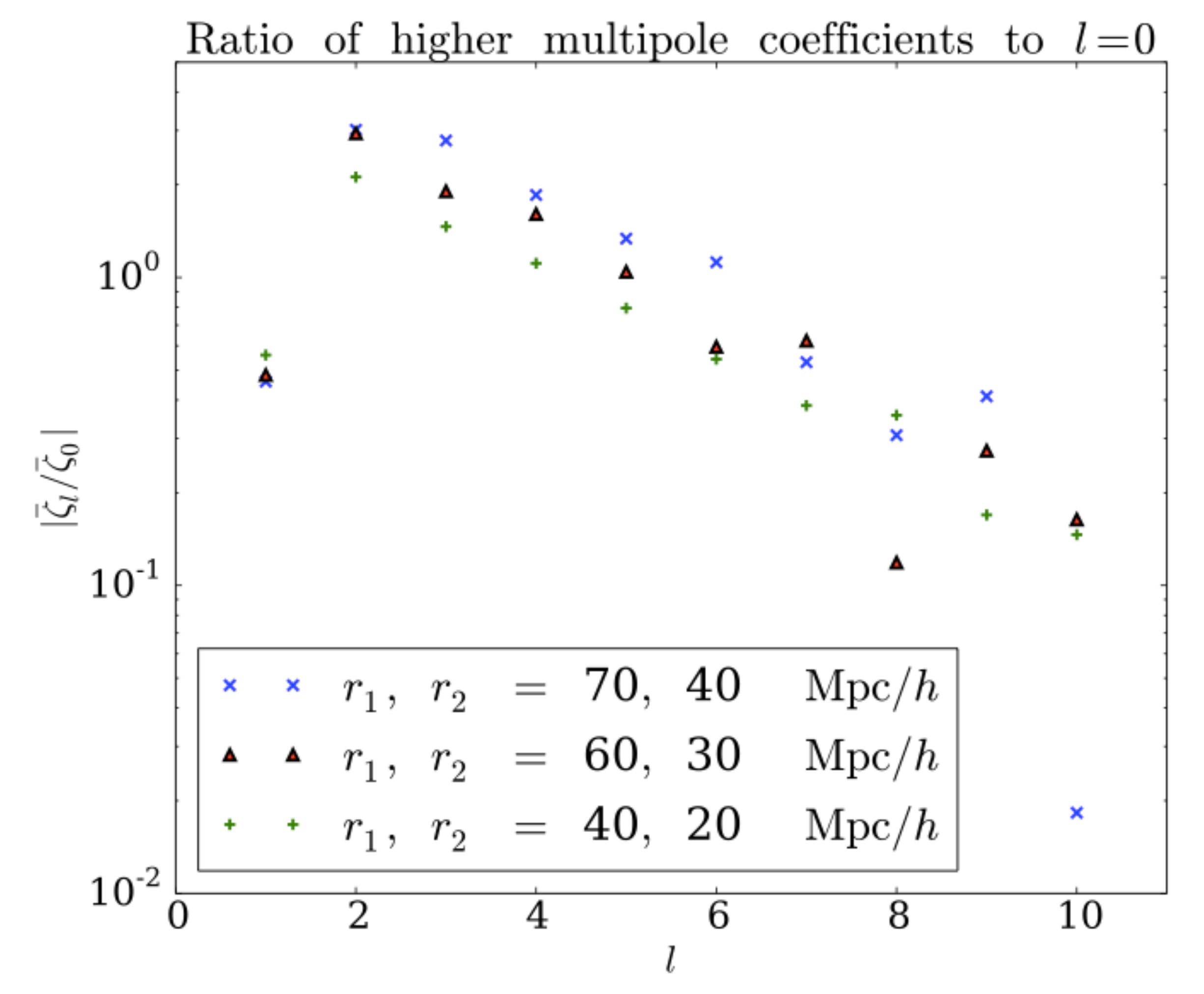}
\caption{This shows the ratios of $l>0$ coefficients to $\bar{\zeta}_0$ for several triangle configurations (again using the LasDamas real space mocks). The decline of the higher multipoles with $l$ indicates that not many multipoles are needed for accurately reconstructing the full 3PCF. This is especially true for the largest scale triangle we show, which is also the least likely to be altered by non-linear effects. The relative magnitudes of the higher multipoles here may seem large when recalling from Figure \ref{fig:recon} that the reconstruction appears well converged by $l=6$; but note that a given multipole's contribution to the reconstruction is roughly its ratio to the sum of all the lower multipoles, not just to $\bar{\zeta}_0$; this reduces the importance of the higher multipoles. Finally, the strength of $l=2$ shows the quadratic or U-shaped behavior of the 3PCF traditionally associated with gravitational growth of structure (see also Figure \ref{fig:recon}). Gravity generates gradients of the density and velocity divergence mostly parallel to the velocity, in turn enhancing roughly collinear structures with $\theta$ near $0$ or $\pi$ (Bernardeau et al. 2002).}
\label{fig:ratios}
\end{figure}

Figures \ref{fig:recon} and \ref{fig:ratios} show that the full 3PCF of the data can be reconstructed well from only a few multipoles.   Figure \ref{fig:recon} reconstructs the 3PCF from coefficients $\zeta_l$, up to and including the $l$ indicated in the legend, for a particular triangle configuration with $r_1=70\;{\rm Mpc}/h$, $r_2=40\;{\rm Mpc}/h$. One recovers an accurate shape versus $\cos\theta_{12}$ even using only multipoles up to $l=6$, and that adding in $l=5-8$ and finally $l=5-10$ changes the shape very little. In Figure \ref{fig:ratios}, we illustrate the same idea for three different triangle configurations: the higher multipoles fall off relative to $\bar{\zeta}_0$, meaning they contribute less to reconstructing the full 3PCF. This plot likely is conservative in that it makes the effect of the higher multipoles appear larger than it is; the plot shows the ratio of each higher multipole to $\bar{\zeta}_0$, but the change in the 3PCF produced by adding in a higher multipole is actually roughly the ratio of the multipole to the sum of {\it all} the lower multipoles.  Since, in detail, Legendre polynomial weights also enter, one might consider an angle-averaged version of this ratio. However since Figure \ref{fig:recon} effectively already shows the unimportance of the highest multipoles, we have in Figure \ref{fig:ratios} just chosen to show $|\bar{\zeta}_l/\bar{\zeta}_0|$ because it offers more granular information.

\subsection{Compressing the data}
\label{subsec:comp_data}
SE15 presented a compression scheme for the multipole moments of the 3PCF. This was designed to avoid the squeezed limit where two galaxies are so nearby that perturbation theory is invalid and also to reduce the dimension of the covariance matrix required for parameter fitting.  This approach integrated each multipole moment over $r_2$ from $r_1/3<r_2<2r_1/3$ at each value of $r_1$.  

In the current work the data is binned coarsely enough in both $r_1$ and $r_2$ that this approach must be adapted slightly.  We simply choose to, for a given bin $r_1$, sum over all bins with $r_2\in S(r_1)$. $S(r_1)$ is the set of all bins in $r_2$ where $r_2$ is greater than $3\Delta r$ and less than $r_1-3\Delta r$.\footnote{Note that if one used different bin sizes $\Delta r$ one might wish to select a different multiple of $\Delta r$ in defining $S$.}  This assures that the minimum value of $r_2$ is $18{\rm \;Mpc}/h$ and that the minimum difference between $r_1$ and $r_2$ is also $18{\rm \;Mpc}/h$, meaning by the Triangle Inequality that $r_3\geq 18{\rm \;Mpc}/h$.  This avoids the squeezed limit while reducing the dimension of the problem. Mathematically, the compression is defined here as

\begin{align}
\bar{\zeta}^c_l(r_1)=\frac{\sum _{r_2\in S(r_1)} \bar{\zeta}_l(r_1,r_2)\Delta V(r_2)}{\sum _{r_2\in S(r_1)}\Delta V(r_2)}
\label{eqn:zeta_comp}
\end{align}
where bar denotes ``binned'', superscript ``c'' denotes ``compression'', and $\bar{\zeta}_l(r_1,r_2)$ is the $l^{th}$ binned 3PCF multipole (see Section \ref{subsec:binning}). $\Delta V(r_2)$ is the volume of bin $r_2$. The denominator is for normalization.

\begin{figure*}
\centering
\includegraphics[scale=.8]{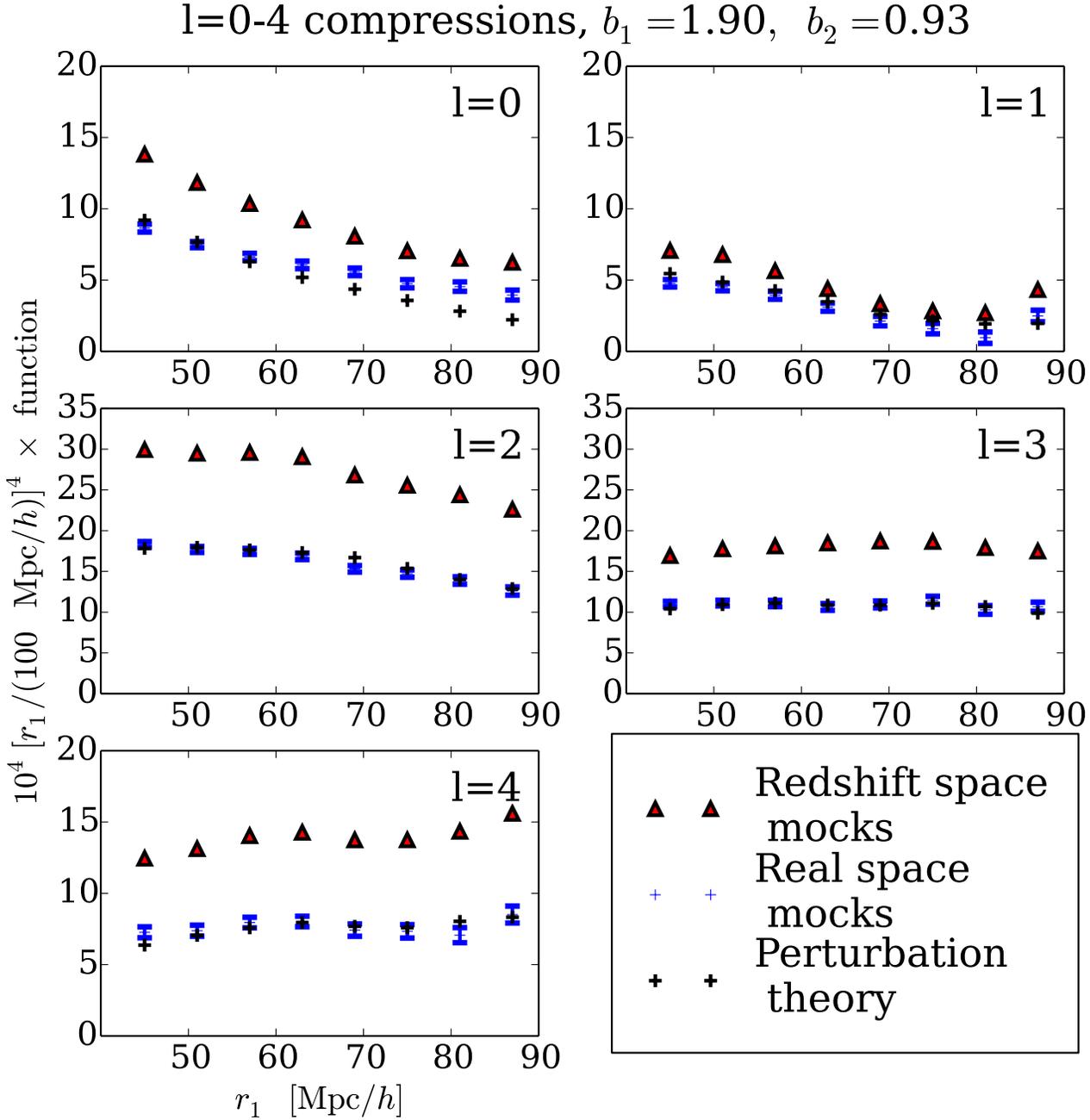}
\caption{The results of compressing the binned multipole moments as described in Section \ref{subsec:comp_data}.  The perturbation theory points are predictions for the 3PCF at lowest (fourth) order, with linear bias $b_1$ and non-linear bias $b_1$ given in the title (see e.g. SE15 equations (1), (2), and Sections 6 and 7; note our non-linear bias does not take a factor of $1/2$ when it multiplies the matter density field's square). Intrinsically the 3PCF is of order $\xi^2$, with $\xi$ the 2-point correlation functon, so on large scales we expect it to be of order $10^{-4}$. We therefore multiply the axis labels by this factor for compactness. Importantly, notice that the redshift space mock results are roughly just a rescaling of the real space mock results by a radius-independent constant that only weakly depends on multipole (see also Figure \ref{fig:redshift_space_rescale}).}
\label{fig:low_ell_comps}
\end{figure*}

\begin{figure*}
\centering
\includegraphics[scale=.8]{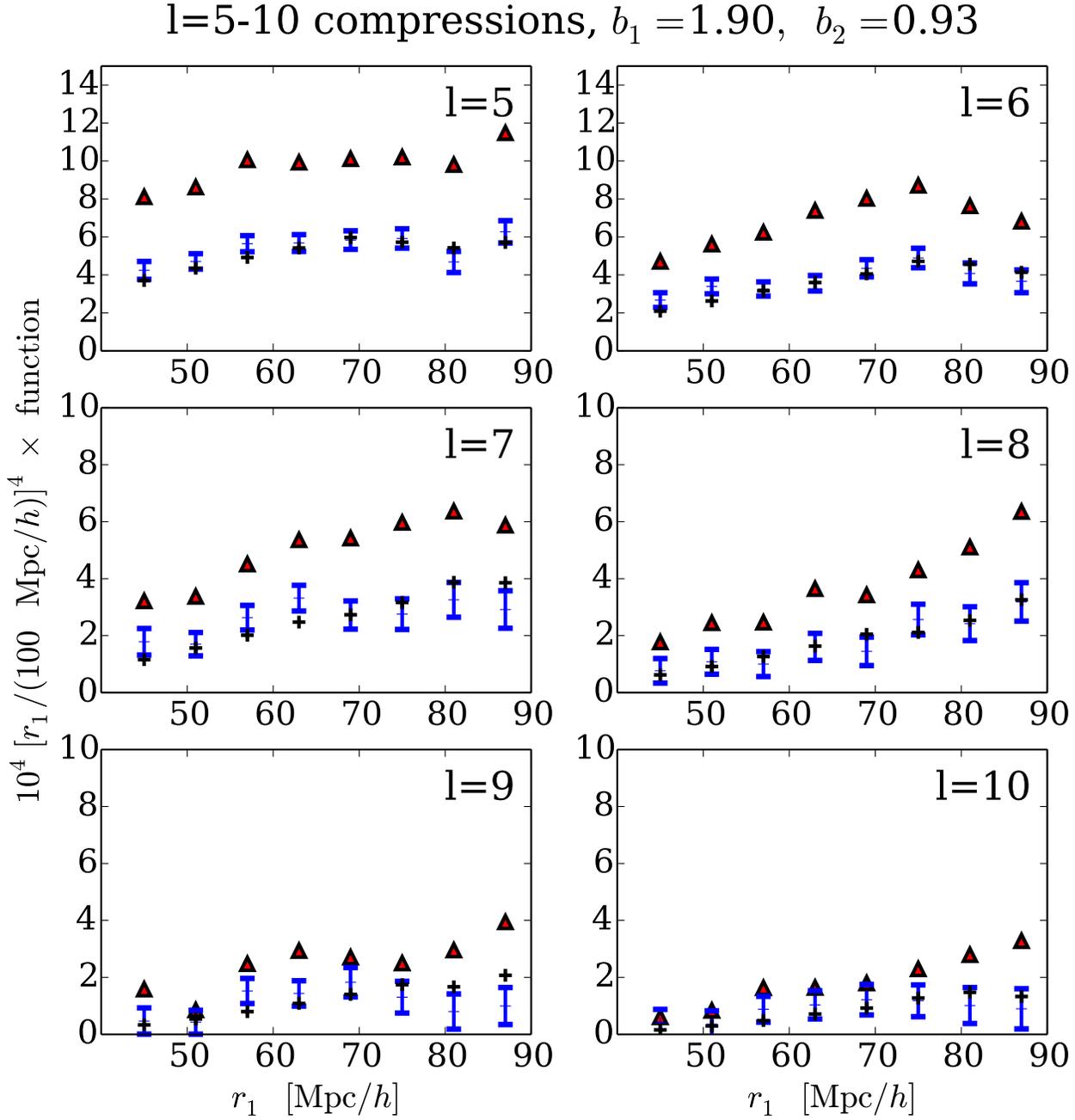}
\caption{Same as Figure \ref{fig:low_ell_comps} but for the higher multipoles, $l=5-10$. Note the lower amplitude of these as compared with especially the lowest multipoles in Figure \ref{fig:low_ell_comps}, and the larger errorbars.}
\label{fig:hi_ell_comps}
\end{figure*}

In Figures \ref{fig:low_ell_comps} and \ref{fig:hi_ell_comps} we show the results of this compression.  We also compressed the leading (fourth) order perturbation theory predictions, calculated as outlined in SE15, and show them for comparison.  The theory requires linear ($b_1$) and non-linear ($b_2$) bias parameters as an input; we use a least-squares fit with points weighted by the inverse compressed variance. This latter is computed from the scatter between mocks and ignores noise in the random catalog used for edge correction, which due to the large number of randoms is negligible.  Our mocks constitute a volume of order 7 times that used for the theoretical covariance calculation here, so, as explained in Figure \ref{fig:covar_diag_grid}, we might expect error bars on the compressions of order $5\%$. This is indeed what we find.  We offer the caveat that a full, rigorously correct fit of theory to observation would require inversion of the full covariance matrix.  We leave this for future work; here our goal is simply to indicate that the results of our algorithm roughly agree with perturbation theory predictions.

Using the simple procedure above, the results are well-fit with $b_1=1.90$ and $b_2=0.93$; note that to compute the theory predictions we matched the LasDamas cosmology.\footnote{LasDamas cosmology given at: \url{http://lss.phy.vanderbilt.edu/lasdamas/simulations.html}; $\sigma_8=0.8$ there is quoted at $z=0$; $\Omega_{\rm m}=0.25,\Omega_{\Lambda}=0.75,n_s=1$. The LasDamas mocks are at $z=0.3$, so when normalizing the power spectrum we should use $\sigma_8(z=0)[D(0.3)/D(0)]$, with $D$ the linear growth factor (e.g. Mo van den Bosch \& White (2010) equations (4.75), (4.76), and (3.77); Carroll et al. (1992)).}
There is some deviation noticeable at large scales in $l=0$ (about $3\sigma$) and $l=1$ (about $2\sigma$) , with nearly all the other multipoles deviating only within the error bars or at most in a few cases just slightly outside them. The larger deviations in $l=0$ and $l=1$ are likely because non-linear corrections to the perturbation theory results cannot be neglected.  In particular, in $l=0$ we expect non-linear evolution might smooth structure on smaller scales, making the slope of the perturbation theory compression shallower and allowing a better global fit to the $l=0$ mock results. 

Importantly, the error bars become much larger for $l\geq 5$ as compared to those for $l < 5$.  This suggests when doing a full parameter fit using the covariance matrix, one might not gain much by including these higher multipoles.  One might choose simply to drop these modes to reduce the dimension of the covariance matrix to be inverted.

\begin{figure}
\includegraphics[scale=0.25]{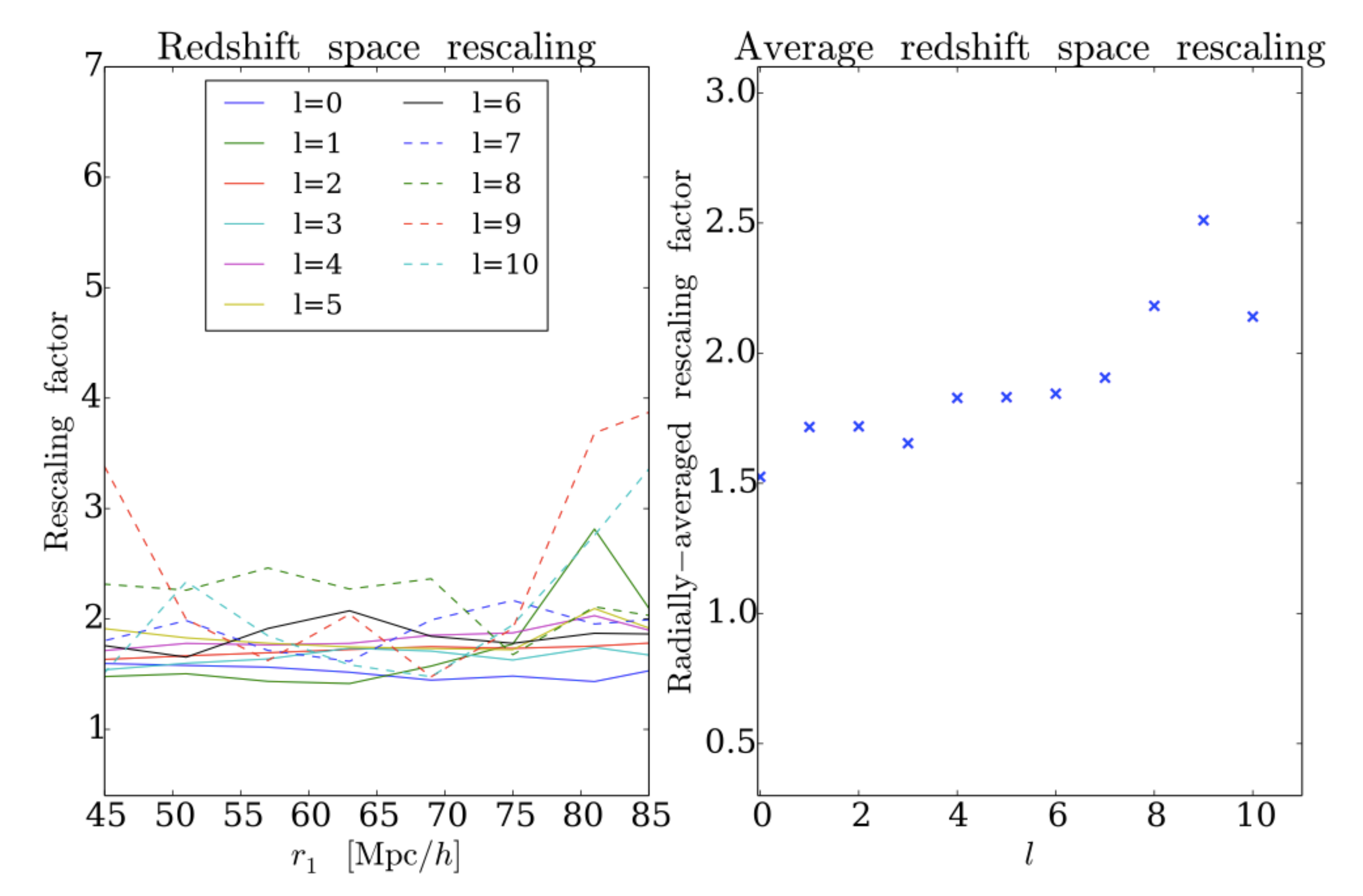}
\caption{The left panel shows the ratio of redshift space to real space results for the LasDamas mocks at each multipole and in each radial bin whose compression is non-zero ($r_1\geq 45\;{\rm Mpc}/h$).  The right panel shows the radial average at each multipole.  While there is some radial scale dependence in the left panel, it is modest for all but $l=8-10$. Thus averaging over the radial dependence does not lose much information for the lower multipoles. The averages (right panel) are similar for all but $l=8-10$, and even these differ by less than a factor of 2 from the averages of the lower multipoles.}
\label{fig:redshift_space_rescale}
\end{figure}

 Figures \ref{fig:low_ell_comps} and \ref{fig:hi_ell_comps} also show that the redshift space results at each multipole appear to be roughly a constant rescaling of the real space results, with a constant that only weakly depends on the multipole.  To illustrate this we show the ratio of redshift space to real space results in each radial bin at each multipole (Figure \ref{fig:redshift_space_rescale}, left panel) and the radially-averaged ratio versus multipole (Figure \ref{fig:redshift_space_rescale}, right panel). More detailed discussion is in the caption to this Figure; the key point is that for $l<8$, there is little radial dependence to the rescaling factor and also little multipole dependence.  Both dependences are more pronounced for $l=8-10$; we suspect this is because these higher multipoles are dominated, even in the compression, by a small subset of relatively squeezed triangles that are more strongly affected by RSD.  This issue might merit further attention in subsequent work.

\subsection{Compressing the covariance}
Given that we now have compressed data, we must also apply our compression scheme as in the previous section to the binned covariance of Section \ref{subsec:binning}.  We denote the compressed, binned covariance as $\overline{{\rm Cov}}_{ll'}^{cc}(r_1;r_1')$, noting that the two superscript ``c''s denote that we compress over $r_2$ and $r_2'$, leaving the quantity a function only of $r_1$  and $r_1'$.

It would be computationally intensive to compute the binned covariance using equation (\ref{eqn:fullcovar_final}) with the $f_{ll}$ and $f_{l_2ll'}$ replaced by equations (\ref{eqn:flbar_2})-(\ref{eqn:jlbar}) and then compress, and a faster approach is available.  This is to compress $\bar{f}_{ll}$ and $\bar{f}_{l_2ll'}$ as necessary first and from them obtain the compressed binned covariance.  $\bar{f}_{ll}$ need only be compressed at most once (if its argument is $r_2$ or $r_2'$), but  $\bar{f}_{l_2ll'}$ may be compressed once or twice depending on if one or both of its arguments have subscript $2$.  We thus define three functions, where ``cc'' again denotes a double compression:

\begin{align}
&\bar{f}_{ll}^c(r;r_1)=\frac{\sum_{r_2\in S(r_1)} \bar{f}_{ll}(r;r_2)\Delta V(r_2)}{\sum_{r_2\in S(r_1)}\Delta V(r_2)},\nonumber\\
&\bar{f}_{l_2ll}^c(r;r_1,r'_1)=\frac{\sum_{r_2'\in S(r_1')} \bar{f}_{l_2ll'}(r;r_1,r_2')\Delta V(r_2')}{\sum_{r_2'\in S(r_1')}\Delta V(r_2')},\nonumber\\
&\bar{f}_{l_2ll}^{cc}(r;r_1,r_1')=\nonumber\\
&\frac{\sum_{r_2\in S(r_1)}\sum_{r_2'\in S(r_1')} \bar{f}_{l_2ll'}(r;r_2,r_2')\Delta V(r_2')\Delta V(r_2)}{\sum_{r_2\in S(r_1)}\sum_{r_2'\in S(r_1')}\Delta V(r_2')\Delta V(r_2)}
\end{align}
Note that in the second line above, the $\bar{f}_{l_2ll'}$ being compressed is a function $r_1$ and $r'_2$, and hence need only be compressed once---it is not compressed over $r_1$. However in the third line above, the $\bar{f}_{l_2ll'}$ being compressed depends on $r_2$ and $r_2'$ and so must be compressed twice. Making these replacements as appropriate in equation (\ref{eqn:fullcovar_final}) yields the compressed, binned covariance. This shows that the framework of compression can be easily generalized to the covariance.

\section{Conclusions}
\label{sec:conclusions}

We have presented a novel algorithm to compute the multipole moments of the 3PCF.  It is especially apt for large cosmological datasets such as SDSS and upcoming surveys like Euclid, Large Synoptic Survey Telescope (LSST), and Dark Energy Spectroscopic Instrument (DESI), which will have tens of millions to billions of objects (Jain et al. 2015).  For these datasets, an approach that scales with $N^3$ would be wholly infeasible.  We have shown that our algorithm scales as $N^2$, handles edge correction easily, and permits computation of the 3PCF of a large dataset quickly even with modest computing resources. We have also computed the covariance matrix of this decomposition in the Gaussian random field limit.  Finally, we have developed the compression scheme first presented in SE15 and shown its application both to the data and to the covariance matrix. This compression scheme offers a compelling way to visualize the results of the algorithm that loses little information, in contrast to the plots of the 3PCF or reduced 3PCF versus opening angle $\theta$ for particular triangle configurations that previous literature supplies.  

The algorithm presented here is unique in that it fundamentally reduces the scaling of the 3PCF measurement to that of the two-point function, while remaining exact in angle. This did not have to be the case.  Formally, for a complete representation of the 3PCF, one needs an infinite number of multipoles $l$.  However, because the physics generating the 3PCF does not have a great deal of angular structure, in practice a finite, modest number of multipoles suffices.   Furthermore, we have shown that in our LasDamas test case, the 3PCF is already well-reconstructed by $l=6$ (Figure \ref{fig:recon}).  Since our algorithm fundamentally requires pair-counting, using a fast Fourier transform (FFT) for this step may in some cases offer an additional acceleration; we present this in Slepian \& Eisenstein 2015c.

Finally, one might worry that jagged survey boundaries could easily introduce high multipoles into the measured 3PCF.  But we have shown that the coefficients required for the edge correction, at least for our LasDamas test case, fall off quickly enough that one need only measure a few multipoles of the randoms for an accurate solution (Figures \ref{fig:f_one} and \ref{fig:Mkl}).

The 3PCF contains important information on the non-Gaussianity of large-scale structure (LSS) due to growth under gravity and also perhaps that remaining from primordial non-Gaussianity. With measurements of only the 2-point function, the amplitude of clustering (e.g. $\sigma_8$) and the linear bias are degenerate. However, the 3PCF is sensitive to a different power of the linear bias than the 2-point function (cube versus square), and so measuring it exposes a raw factor of the bias and helps break this degeneracy.  As for primordial non-Gaussianity,  while the CMB has been the dominant constraint up to now (see Ade et al. (2015)), it is expected that even maximally improved CMB measurements can only enhance the CMB constraint by a factor of a few.  Thus LSS will become a vital complementary probe.  Generically, inflation must couple to ordinary matter so as to produce it during reheating, and this coupling produces some level of non-Gaussianity (Desjacques \& Seljak 2010, for a recent review).  Thus the 3PCF can be used to probe the dynamics of inflation in principle---and perhaps, soon, in practice.  

The 3PCF also contains information on redshift space distortions. It should be emphasized that in the current work, the multipole moments are averaged over rotations of the triangle configurations, and so we lose any information about anisotropy.  However, our algorithm can easily be adapted to retain the full, unaveraged information ($a_{lm}$s) around each possible origin.  This would allow tracing the anisotropy RSD induce.  Preliminary calculations indicate that RSD introduce couplings between multipoles $l\neq l'$ which are absent without RSD.  These couplings have selection rules due to the underlying symmetries under rotation about the line of sight and parity flips. There will also be $m$ dependence induced by the preferred direction defined by the line-of-sight. We therefore expect that the off-diagonal terms in a tensor of spherical harmonic coefficients will have structure that can be used to probe RSD-induced anisotropies.  It is already important that the spherical harmonics, with the introduction of $l'\neq l$ and $m$, offer a natural 5-D basis for redshift-space measurements. Work on these questions from both analytic and numerical perspectives is underway.

We plan to apply our algorithm and analysis approach to SDSS DR12, with the goals of assessing the presence of BAO features, measuring the linear and non-linear bias, and constraining the baryon-dark matter relative velocity (Tseliakhovich \& Hirata 2010; Yoo, Dalal \& Seljak 2011; Yoo \& Seljak 2013; SE15). Previous literature has found a BAO feature in the reduced 3PCF $\zeta/\xi^2$, with $\xi$ the 2PCF (Gazta\~{n}aga et al. 2009).  That work used only one triangle configuration, $r_1=33\pm5.5{\rm\; Mpc}/h$ and $r_2=88\pm5.5{\rm\; Mpc}/h$.  With the additional signal-to-noise the large number of galaxies in SDSS DR12 offers, as well as our algorithm's ability to consider all triangle configurations quickly, this problem is ripe for revisiting.  Furthermore, SE15 suggests that the multipole decomposition clearly isolates a strong BAO feature, especially in the $l=1$ multipole. This is also a particularly informative multipole for the relative velocity, further discussed in SE15. That work additionally shows that, in principle, the multipole decomposition can clearly separate the effects of linear and non-linear bias---significant because, as noted above, the 3PCF has traditionally been an important tool for constraining these parameters. Finally, the significant speed advantage of our algorithm will permit much finer and much faster calibration of any 3PCF measurements against large cosmological simulations.  Such improved calibration should greatly enhance the leverage of the 3PCF as a fundamental probe of large-scale structure.

\section*{Acknowledgments}
ZS thanks Simeon Bird, Doug Finkbeiner, Lehman Garrison, JR Gott III, Robert Marsland, Philip Mocz, Cameron McBride, Stephen Portillo, David Spergel, and Yucong Zhu for useful discussions.  We thank the anonymous referee for several helpful suggestions as well. This material is based upon work supported by the National Science Foundation Graduate Research Fellowship under Grant No. DGE-1144152.
\section*{References}

\hangindent=1.5em
\hangafter=1
\noindent Adams J. C., 1878, Proc. R. Soc., 27, 63

\hangindent=1.5em
\hangafter=1
\noindent Ade P.A.R. et al., 2015, preprint (arXiv:1502.01592)

\hangindent=1.5em
\hangafter=1
\noindent Arfken G.B., Weber H.J. \& Harris F.E., 2013, Mathematical Methods for Physicists: Academic Press, Waltham, MA

\hangindent=1.5em
\hangafter=1
\noindent Bernardeau F., Colombi S., Gazta\~{n}aga E., Scoccimarro R., 2002, Phys. Rep., 367, 1

\hangindent=1.5em
\hangafter=1
\noindent Chen G. \& Szapudi I., 2005, ApJ, 635:743-749

\hangindent=1.5em
\hangafter=1
\noindent Desjacques V. \& Seljak U., 2010, Classical and Quantum Gravity, vol. 27, issue 12

\hangindent=1.5em
\hangafter=1
\noindent Fabrikant V.I., 2013, Quarterly of Applied Mathematics, vol. LXXI, 3, 573-581

\hangindent=1.5em
\hangafter=1
\noindent Ferrers N.M., 1877, An elementary treatise on Spherical Harmonics and Subjects Connected
with them: Macmillan, London

\hangindent=1.5em
\hangafter=1
\noindent Frieman J.A. \& Gazta\~{n}aga E., 1999, ApJ, 521, L83-86.

\hangindent=1.5em
\hangafter=1
\noindent Gardner J. P., Connolly A. \& McBride C., 2007, in ASP Conf Ser. 376, Astronomical Data Analysis Software and Systems XVI, ed. R. A. Shaw, F. Hill, \& D. J. Bell (San Francisco, CA: ASP), 69

\hangindent=1.5em
\hangafter=1
\noindent Gazta\~{n}aga E. \& Frieman J.A., 1994, ApJ, 437, L13.

\hangindent=1.5em
\hangafter=1
\noindent Gazta\~{n}aga E., Cabr\'e A., Castander F., Crocce M. \& Fosalba P., 2009, MNRAS 399, 2, 801-811

\hangindent=1.5em
\hangafter=1
\noindent Gradshteyn I.S. \& Ryzhik I.M., 2007, Table of Integrals, Series, and Products, ed. A. Jeffrey \& D. Zwillinger (Amsterdam: Academic Press)

\hangindent=1.5em
\hangafter=1
\noindent Gray A. G., Moore A. W., Nichol R. C., Connolly A. J., Genovese C., \& Wasserman L. 2004, in ASP Conf. Ser. 314: Astronomical Data Analysis Software and Systems (ADASS) XIII, 249

\hangindent=1.5em
\hangafter=1
\noindent Goroff M.H., Grinstein B,, Rey S.-J. \& Wise M.B., 1986, ApJ, 311, 6-14

\hangindent=1.5em
\hangafter=1
\noindent Guo H. et al., 2015, MNRAS 449, 1, L95-L99

\hangindent=1.5em
\hangafter=1
\noindent Jain B. \& Bertschinger E., 1994, ApJ, 431: 495-505

\hangindent=1.5em
\hangafter=1
\noindent Jain B. et al., 2015, preprint (arXiv:1501.07897v2)

\hangindent=1.5em
\hangafter=1
\noindent Jing Y. P. \& B\"{o}rner G., 2004, ApJ, 607, 140

\hangindent=1.5em
\hangafter=1
\noindent Kayo I. et al., 2004, PASJ, 56, 415

\hangindent=1.5em
\hangafter=1
\noindent Landy S.D. \& Szalay A.S., 1993, ApJ, 412, 1 

\hangindent=1.5em
\hangafter=1
\noindent Lewis A., 2000, ApJ, 538, 473

\hangindent=1.5em
\hangafter=1
\noindent March W.B., 2013, PhD Thesis, Multi Tree Algorithms for Computational Statistics and Physics

\hangindent=1.5em
\hangafter=1
\noindent McBride C., Connolly A. J., Gardner J. P., Scranton R., Newman J., Scoccimarro R., Zehavi I., Schneider D. P., 2011a, ApJ, 726, 13

\hangindent=1.5em
\hangafter=1
\noindent McBride K., Connolly A. J., Gardner J. P., Scranton R., Scoccimarro R., Berlind A., Marin F., Schneider D. P., 2011b, ApJ, 739, 85

\hangindent=1.5em
\hangafter=1
\noindent Mehrem R., 2011, Journal of Applied Mathematics and Computation 217 5360-5365

\hangindent=1.5em
\hangafter=1
\noindent Moore A. W. et al., 2001, in Mining the Sky, ed. A. J. Banday, S. Zaroubi, \& M. Bartelmann (Berlin: Springer), 71

\hangindent=1.5em
\hangafter=1
\noindent Neumann F.E., 1878, Beitrage zur Theorie der Kugelfunctionen, Teubner: Leipzig

\hangindent=1.5em
\hangafter=1
\noindent NIST Digital Library of Mathematical Functions (DLMF).\\ 
http://dlmf.nist.gov/, release 1.0.10 of 2015-08-07. \\
Online companion to W. J. Olver, D. W. Lozier, R. F. Boisvert, and C. W. Clark, eds. 
NIST Handbook of Mathematical Functions. 
Cambridge University Press, New York, NY, 2010.

\hangindent=1.5em
\hangafter=1
\noindent Nichol R. C. et al., 2006, MNRAS, 368, 1507

\hangindent=1.5em
\hangafter=1
\noindent Padmanabhan, N., White, M., \& Eisenstein, D.J., 2007, \mnras, 376, 1702

\hangindent=1.5em
\hangafter=1
\noindent Pan J., Szapudi I., 2005, MNRAS, 362, 4, 1363

\hangindent=1.5em
\hangafter=1
\noindent Park S.B. \& Kim J.-H., 2006, J. Appl. Math. \& Computing 20, 1-2, 623-635

\hangindent=1.5em
\hangafter=1
\noindent Peebles P. J. E. \& Groth E. J., 1975, ApJ, 196, 1

\hangindent=1.5em
\hangafter=1
\noindent Slepian Z. \& Eisenstein D.J., 2015a, MNRAS 448, 9-26

\hangindent=1.5em
\hangafter=1
\noindent Slepian Z. \& Eisenstein D.J., 2015c, MNRAS in press, preprint available at arXiv:1506.04746

\hangindent=1.5em
\hangafter=1
\noindent Szapudi I., Szalay A., 1998, ApJ, 494, L41

\hangindent=1.5em
\hangafter=1
\noindent Szapudi I., Szalay A., 1998, ApJ, 494, L41

\hangindent=1.5em
\hangafter=1
\noindent Szapudi I., 2001, proc. of ``The Onset of Nonlinearity in Cosmology'', ed. J.N. Fry, J.R. Buchler \& H. Kandrug, in Annals of the New York Academy of Sciences, vol. 927. 

\hangindent=1.5em
\hangafter=1
\noindent Szapudi I., 2004, ApJ, 605, L89

\hangindent=1.5em
\hangafter=1
\noindent Szapudi I., 2005, chapter in ``Data Analysis in Cosmology'', ed. V.J. Martinez, E. Martinez-Gonzalez, M.J. Pons-Borderia \& E. Saar, Springer-Verlag Lecture Notes in Physics.

\hangindent=1.5em
\hangafter=1
\noindent Tseliakhovich D. \& Hirata C.M., 2010, PRD 82, 083520

\hangindent=1.5em
\hangafter=1
\noindent Yoo J., Dalal N. \& Seljak U,, 2011, JCAP 1107:018

\hangindent=1.5em
\hangafter=1
\noindent Yoo J. \& Seljak U., 2013, PRD 88, 10

\hangindent=1.5em
\hangafter=1
\noindent Zhang L.L. \& Pen U.-L., 2005, New Astronomy, 10, 7, 569-590.

\end{document}